\documentclass[prl,twocolumn,superscriptaddress]{revtex4-1}
\usepackage[utf8]{inputenc}
\usepackage[american,british]{babel}
\usepackage[T1]{fontenc}
\usepackage[pdftex]{graphicx}  
\usepackage{graphicx, xcolor}
\usepackage{dcolumn}
\usepackage{bm}
\usepackage{amsmath,amsthm,amssymb,mathtools,multirow}
\usepackage{color}
\usepackage{verbatim}
\usepackage{ulem}

\definecolor{darkGreen}{RGB}{0,110,0}
\definecolor{darkBlue}{RGB}{0,0,130}
\usepackage[colorlinks,citecolor=darkGreen,linkcolor=darkBlue,urlcolor=blue,hyperindex]{hyperref}

\newcommand{\bra}[1]{\left\langle #1 \right|}
\newcommand{\ket}[1]{\left| #1 \right\rangle}
\newcommand{\braket}[2]{\left\langle #1 \middle| #2 \right\rangle}
\newcommand{\brakett}[3]{ \langle #1 | #2 | #3 \rangle}
\newcommand{\ketbra}[2]{\left|#1\middle\rangle\middle\langle#2\right|}

\usepackage[capitalize]{cleveref} 

\newcommand{\titleinfo}{Unraveling PXP Many-Body Scars through Floquet Dynamics}

\begin{document}
\title{\titleinfo}
\author{Giuliano Giudici}
\affiliation{Institute for Theoretical Physics, University of Innsbruck, Innsbruck 6020, Austria}
\affiliation{Institute for Quantum Optics and Quantum Information,Austrian Academy of Sciences, Innsbruck 6020, Austria}
\affiliation{planqc GmbH, Garching 85748, Germany}
\author{Federica Maria Surace}
\affiliation{Department of Physics and Institute for Quantum Information and Matter,
    California Institute of Technology, Pasadena, California 91125, USA}
\author{Hannes Pichler}
\affiliation{Institute for Theoretical Physics, University of Innsbruck, Innsbruck 6020, Austria}
\affiliation{Institute for Quantum Optics and Quantum Information,Austrian Academy of Sciences, Innsbruck 6020, Austria}

\date{\today}

\begin{abstract} 
Quantum scars are special eigenstates of many-body systems that evade thermalization. They were first discovered in the PXP model, a well-known effective description of Rydberg atom arrays. Despite significant theoretical efforts, the fundamental origin of PXP scars remains elusive. 
By investigating the discretized dynamics of the PXP model as a function of the Trotter step $\tau$, we uncover a remarkable correspondence between the zero- and two-particle eigenstates of the integrable Floquet-PXP cellular automaton at $\tau=\pi/2$ and the PXP many-body scars of the time-continuous limit. Specifically, we demonstrate that PXP scars are adiabatically connected to the eigenstates of the $\tau=\pi/2$ Floquet operator. Building on this result, we propose a protocol for achieving high-fidelity preparation of PXP scars in Rydberg atom experiments.
\end{abstract} 

\maketitle

\paragraph{Introduction. --} 
Quantum scars in many-body systems were first discovered following the unexpected dynamics experimentally observed in a one-dimensional Rydberg atom array \cite{Bernien2017}. Even though the Hamiltonian describing this system is non-integrable, local observables after a quench do not reach thermal equilibrium but instead display persistent oscillations with a period of unexplained nature. Crucially, the appearance of this phenomenon hinges on the initial conditions.
This observation led to the understanding that the spectrum of the Rydberg atoms Hamiltonian contains a peculiar set of eigenstates \cite{Turner2018}. These states account for most of the overlap of the non-thermalizing initial condition, and their energy spacing dictates the period of the oscillations. Borrowing from the literature on single-particle quantum mechanics \cite{Heller1984}, they are called quantum many-body scars. In that context, some eigenstates exhibit scars that trace the existence of unstable periodic orbits in the corresponding classical system. Scar eigenstates possess wavefunctions localized around these fine-tuned trajectories, reflecting the influence of classical dynamics on their quantum behavior.
Despite this suggestive analogy, a rigorous relationship between classical periodic orbits and quantum scars in many-body systems remains far from settled. In light of this incomplete theoretical picture, several attempts have been made to explain the origin of scarred eigenstates in the Rydberg atom model, based on diverse theoretical frameworks ranging from the time-dependent variational principle \cite{Ho2019}, SU(2) algebras \cite{Choi2019}, emerging integrability \cite{Khemani2019}, and many others \cite{lin2019,pimagnons,bull2020,turner2021,muller2023,desaules2022,Serbyn2021,chandran2023,nishad2021}.\\
\begin{figure}
 \includegraphics[scale=0.52]{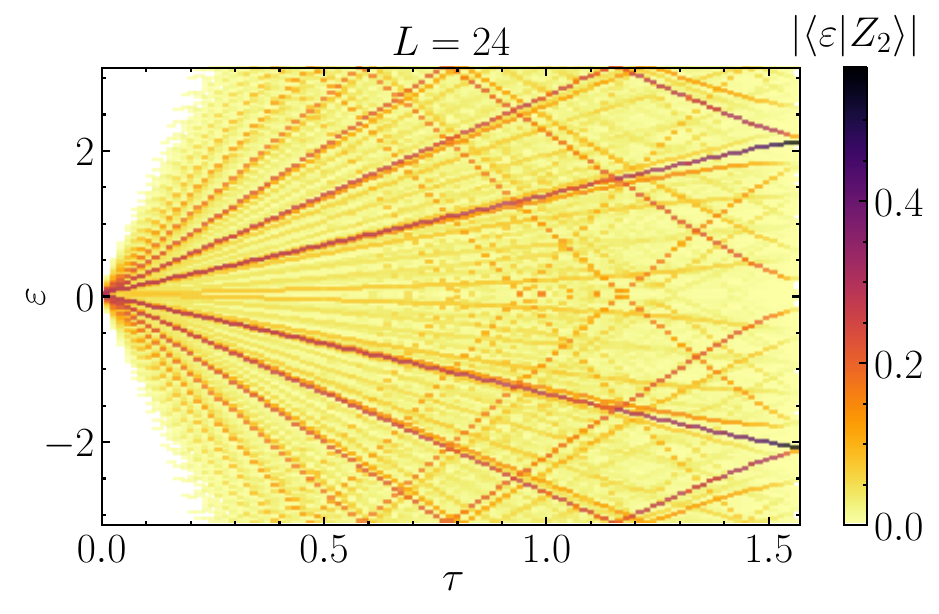}
  \vspace*{-3mm}
\caption{Overlap between the eigenstates at quasienergy $\varepsilon$ of the Floquet operator \cref{eq:floquet} and the $Z_2$ state $\ket{01}^{\otimes L/2 }$ in the symmetry sector $T^2 = T R = 1$, as a function of the Trotter step $\tau$ and for $L=24$ lattice sites. The scars of the PXP Hamiltonian ($\tau\to0$) are smoothly connected to the eigenstates of the classical cellular automaton CA ($\tau=\pi/2$). }
 \label{fig1}
\end{figure}
\indent
In this work, we connect the many-body scars to few-particle eigenstates of an integrable model that arises from a time-discretization of the Rydberg atom dynamics. We focus on the PXP model, an effective description of the Rydberg atoms system where the strong repulsive interaction between Rydberg states is modeled by a constraint that prohibits the simultaneous excitation of two nearby atoms \cite{Fendley2004,lesanoski2011}. By discretizing the PXP dynamics and using the Trotter step $\tau$ as a parameter, we interpolate between the time-continuous limit $\tau \to 0$ and the integrable PXP cellular automaton (CA) $\tau = \pi/2$ \cite{Prosen2020}. Through exact diagonalization of the Floquet operator, we show that high-temperature~\cite{footnote3} PXP quantum scars directly flow into eigenstates of the integrable model with zero or two quasiparticles. We find that the quasienergies $\varepsilon(\tau)$ of the interpolating eigenstates are almost exactly linear in $\tau$ well beyond the perturbative regime $\tau \ll 1$. The slope of these lines determines the energy splitting between the $\tau \to 0$ scars to be $\Delta E= 4/3$, in agreement with finite-size extrapolation \cite{Turner2018b}. We show that signatures of the integrable dynamics of the $\tau = \pi/2$ model persist up to the time-continuous limit, explaining the previously observed ballistic propagation of defects on top of charge-density-wave states \cite{Surace2020}. Finally, we make use of this correspondence to engineer dynamical protocols for the preparation of PXP many-body scars with impressively large fidelities, starting from the simple eigenstates of the $\tau = \pi/2$ integrable Floquet operator. We conclude by discussing connections to previous works and extensions to more than one dimension.
\vspace*{2mm}
\paragraph{PXP model and integrable celullar automaton. --} The PXP model on a chain describes a system of $L$ qubits $\{ \ket{0}, \ket{1} \}$ and its Hamiltonian reads
\begin{equation}
    H = \sum_{j=1}^L P_{j-1} X_j P_{j+1},
    \label{eq:PXP}
\end{equation}
where $X_j = \ket{0}_j\!\bra{1} + \ket{1}_j\!\bra{0}$, $P_j = \ket{0}_j\!\bra{0}$, and periodic boundary conditions are assumed. The Hamiltonian \cref{eq:PXP} commutes with the operators $n_j n_{j+1}$, where $n_j = 1 - P_j$, and, in the Hilbert space sector where $n_j n_{j+1} = 0$ for all $j$, it provides an effective description of a one-dimensional array of Rydberg atoms subject to a strongly repulsive van der Waals force \cite{Fendley2004,lesanoski2011}. In this sector, the PXP Hamiltonian hosts $L+1$ scar eigenstates that are approximately equally spaced in energy and which have large overlaps with the state $\ket{Z_2} = \ket{ 0 1 }^{\otimes L/2}$ (and its translation). The short-time dynamics that emerges from this initial state takes place almost entirely in this $(L+1)$-dimensional subspace. This fact, combined with the almost uniform scars energy spacing $\Delta E$, leads to periodic revivals of the $Z_2$ state with period $T = 2 \pi / \Delta E$. The existence of these peculiar eigenstates was put forward in Ref.~\cite{Turner2018} as the phenomenological explanation for the anomalous dynamics of the $Z_2$ state, however, the reason why they arise in the spectrum of the PXP Hamiltonian remains unclear. Here, we argue that they are connected to some eigenstates of an integrable model resulting from a time-discretization of the PXP dynamics.\\
\indent
We consider the Floquet operator
\begin{equation}
    U_\tau = U^e_\tau \, U^o_\tau = e^{-i H_e \tau } e^{-i H_o \tau },
    \label{eq:floquet}
\end{equation}
where $H_o$ ($H_e$) is the Hamiltonian \cref{eq:PXP} with the sum restricted to odd (even) sites of the chain. We refer to  $\tau$ as the Trotter step size. For $\tau = \pi/2$ this time-discrete model is equivalent to an integrable classical CA and the spectrum can be found analytically from the orbits of classical configurations \cite{SupMat}. In particular, the state $\ket{0}^{\otimes L}$, and the two $Z_2$ states $\ket{01}^{\otimes L/2 }$ and $\ket{10}^{\otimes L/2}$ (the three {\it vacua}) exhibit period-3 oscillations under $U_{\pi/2}$. As an immediate consequence, we get the following eigenstates of $U_{\pi/2}$ with eigenvalues $\lambda = e^{i \varepsilon_0}$
\begin{equation}
\ket{\varepsilon_0} = \frac{\ket{0}^{\otimes L}+e^{i(\varepsilon_0+\frac{\pi}{4} L)}\ket{01}^{\otimes \frac{L}{2}}+e^{-i(\varepsilon_0+\frac{\pi}{4} L)} \ket{10}^{\otimes \frac{L}{2}}}{\sqrt{3}},
\label{eq:vacua}
\end{equation}
where $\varepsilon_0 =  0 , \pm 2 \pi / 3$.
The other eigenstates of $U_{\pi/2}$
can be expressed in terms of chiral quasiparticles that interpolate between the three vacua.  The numbers of right ($+$) and left ($-$) moving quasiparticles are the conserved charges $Q^\pm = \sum_j q^\pm_{j,..,j+3} $, where $q^+$ is a diagonal operator with support on four nearby sites and non-zero matrix elements $\brakett{1001}{q_{j,..}^+}{1001} = 1$ for $j$ even, $\brakett{0001}{q_{j,..}^+}{0001} = \brakett{1000}{q_{j,..}^+}{1000} = 1$ for $j$ odd, and $q^- = T q^+$, with $T$ the one-site translation operator \cite{Prosen2020}. Periodic boundary conditions impose $Q^+ - Q^- = 0 \, (\mathrm{mod}\, 3)$. 
 By applying the Floquet operator \cref{eq:floquet} to an initial state with localized quasiparticles, one sees that left and right movers propagate ballistically by two lattice spacings every three time steps \cite{Prosen2020,iadecola_ca,cesa2023}, i.e., their velocity in real time is $v = 2/3 \tau = 4/3 \pi $.

\begin{figure}
 \includegraphics[scale=0.52]{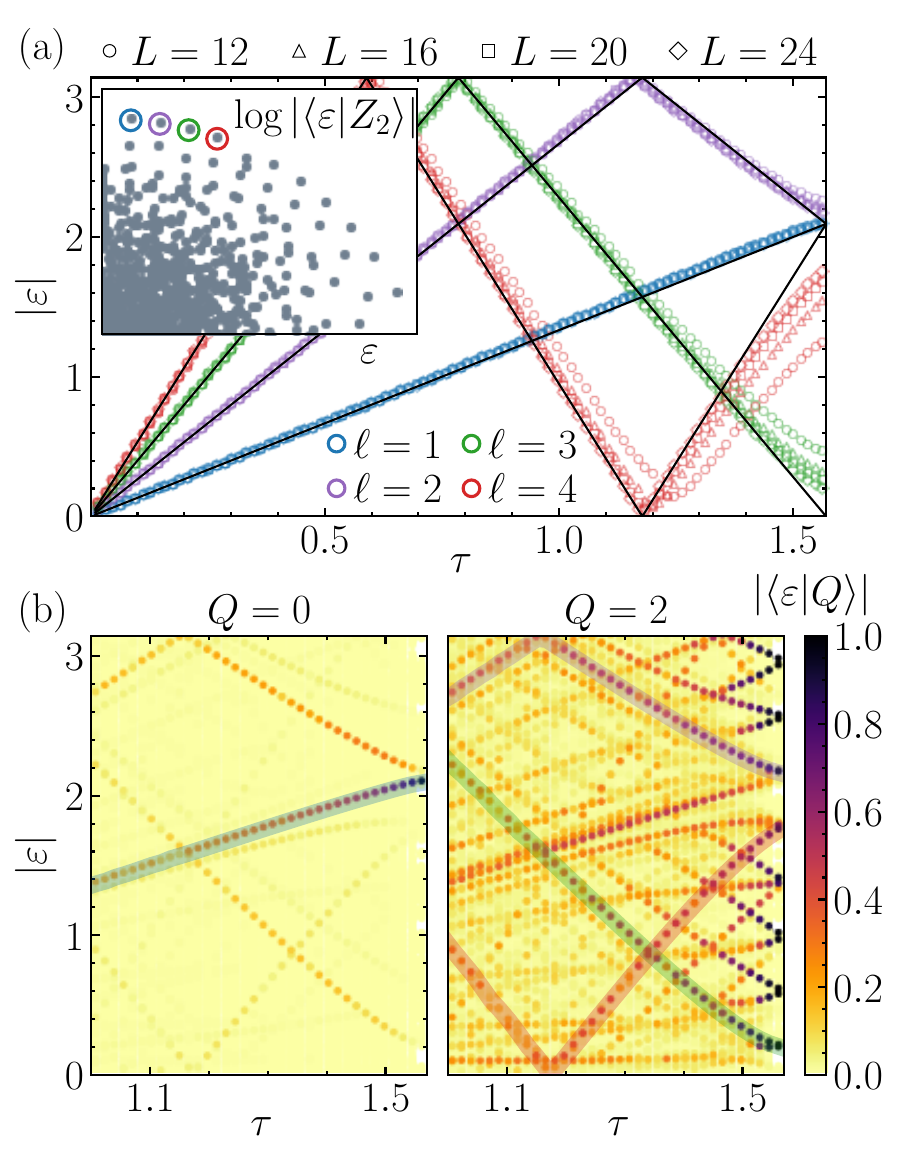}
 \vspace*{-3mm}
\caption{(a) Quasienergies $\varepsilon_\ell$ obtained by adiabatic continuation (see main text) of the $\tau = 0$ positive-energy scars (depicted in the inset), as functions of $\tau$ and for different system sizes. The index $\ell$ labels the time-continuous positive-energy PXP scars starting from $\varepsilon=0$. (b) Overlaps of the $Q=0$ and $Q=2$ subspaces with the eigenstates of the Floquet operator \cref{eq:floquet} for $\tau \in [1,\pi/2]$. The colored thick lines correspond to the adiabatic eigenstates depicted in panel (a). }
\label{fig2}
\end{figure}
\vspace*{1mm}
\paragraph{From integrable eigenstates to PXP scars. --} We now demonstrate that the integrable eigenstates of the unitary operator $U_{\pi/2}$ \cref{eq:floquet} and the many-body scars of the PXP Hamiltonian \cref{eq:PXP}, i.e., $\lim_{\tau \to 0 } i \log U_\tau/\tau$, are smoothly connected as $\tau$ is varied between $0$ and $\pi/2$. To this end, we diagonalize $U_\tau$ in the sector invariant under all its symmetries, namely the two-site translation $T^2$ and one-site translation combined with reflection $T R$. For each $\tau$ we compute the overlap between the eigenstates $\ket{\varepsilon}$ and a $Z_2$ state, to produce the density plot in \cref{fig1}. In the continuous-time limit, the eigenstates with the largest $Z_2$ overlap in a fixed energy window are exactly the many-body scars. It is apparent that not only does such a set of states exist for all $\tau$, but their quasienergies change linearly with $\tau$, even for Trotter steps sizes far beyond the perturbative regime $\tau \ll 1$, and reach values close to the vacua quasienergies $\varepsilon_0$ for $\tau \to \pi/2$.\\
\indent
We make this connection more precise as follows. For a given scar eigenstate of the PXP Hamiltonian \cref{eq:PXP} with $\varepsilon > 0$, we successively find the scar at Trotter step $\tau_{k+1} = \tau_k + d \tau $ by selecting the eigenstate of $U_{\tau_{k+1}}$ with largest overlap with the scar eigenstate of $U_{\tau_k}$. In this way, we can draw the curves $\varepsilon_\ell(\tau)$ depicted in \cref{fig2}(a), which represent the scar eigenvalues as a function of $\tau$ \cite{footnote4}. Here $\ell = 1,2,..$ labels the scars of the PXP Hamiltonian starting from zero energy (cf. \cref{fig2}(a)). Remarkably, we obtain smooth curves that approach, with increasing $L$, the straight lines $(4/3) \tau \ell$ (black solid lines in \cref{fig2}(a)). The slope of these lines is fixed by the assumption, motivated below, $\varepsilon_\ell(\pi/2) = 0, \pm 2 \pi/3$. This fact implies that the energy gap between the $\tau\rightarrow 0$ scars is $(\epsilon_{\ell+1} (\tau)-\epsilon_{\ell} (\tau))/\tau =4/3$. Therefore, the period of the $Z_2$ oscillations in the PXP model \cref{eq:PXP} is ultimately given by the period-3 oscillations of this state under $U_{\pi/2}$ in real time, namely $T_{Z_2} = 3 \tau = 3 \pi / 2 \simeq 4.71 $. 
For $\ell > 4$ we could not ``adiabatically follow'' the scar eigenstates in this fashion.\\
\indent
We note that the linearity of $\varepsilon_\ell (\tau)$ for small $\tau$ can be partially explained by observing that the $O(\tau^2)$ correction vanishes on all eigenvalues of $U_{\tau \to 0}$ \cite{footnote1}.
However, the remarkable fact is that the scar eigenvalues do not hybridize with the rest of the spectrum all the way to $\tau = \pi/2$. In particular, the $\tau \to 0$ scar with $\ell = 1$ directly flows into the vacuum state \cref{eq:vacua} of the integrable Floquet operator $U_{\pi/2}$ with quasienergy $\varepsilon = 2 \pi/3$ and $Q=Q^+ + Q^- = 0$. As we show in \cref{fig2}(b), the remaining three positive-energy scars that we can follow with the procedure outlined above approach integrable eigenstates with two quasiparticles ($Q=2$) for $\tau \to \pi/2$. The quasienergies $\varepsilon$ of these eigenstates converge to one of the vacua quasienergies $\varepsilon_0$ for $L\to\infty$ and $\tau=\pi/2$, in particular $| \varepsilon - \varepsilon_0 | = O( 1/L )$ \cite{SupMat}. It is thus natural to expect that $\varepsilon_\ell (\tau) \to 0,\pm 2 \pi/3$ for $\tau \to \pi/2$, in the thermodynamic limit.\\
\indent
We stress that the eigenvalues $\varepsilon_\ell$ are not perfect lines even for $L \to \infty$ \cite{SupMat}, indicating that the correspondence between integrable eigenstates and time-continuous scars is not exact. This is not unexpected, since it is known that scars in the PXP model \cref{eq:PXP} are only approximately decoupled from the rest of the spectrum \cite{Choi2019}. 
Nevertheless, as we show below, other signatures of the integrable Floquet dynamics can be detected in the PXP Hamiltonian dynamics. 

\begin{figure}
\includegraphics[scale=0.51]{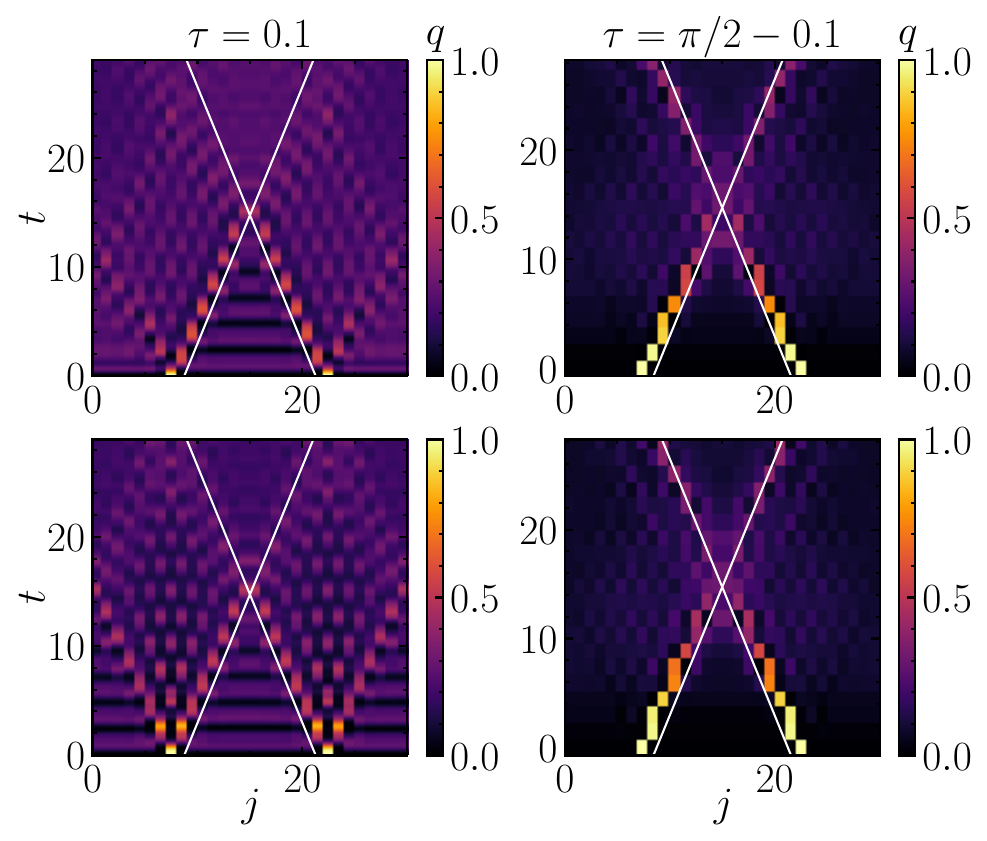}
 \vspace*{-2mm}
\caption{Quasiparticles scattering approaching the time-continuous limit (left) and the integrable point (right). In the upper (lower) panels the quasiparticles are kinks between $\ket{0}^{\otimes L}$ and $\ket{01}^{\otimes L/2}$ ($\ket{10}^{\otimes L/2}$ and $\ket{01}^{\otimes L/2}$). The solid white lines are $x = x_0 \pm v t$ with $v=4/3 \pi$, the kinks are initially at a distance of 14 sites and $L=30$.}
\label{fig3}
\end{figure}

\paragraph{Integrable dynamics in the PXP model. --} 
Chiral quasiparticles of the integrable model $\tau = \pi/2$ propagate ballistically and interact with each other retaining their identity after scattering \cite{Prosen2020}. This can be observed by preparing an initial state with localized kinks between the three vacua $\ket{0}^{\otimes L },\ket{01}^{\otimes L/2},\ket{10}^{\otimes L/2}$ and letting it evolve with $U_{\pi/2}$ \cref{eq:floquet}.
Here, we show that signatures of this integrable dynamics can be observed for all Trotter steps $\tau$.
In \cref{fig3} we plot the quasiparticles density $q = q^+ + q^-$ time evolution starting from initial states with two localized kinks initially apart and propagating towards each other. For $\tau = \pi/2 -  0.1$, i.e., close to the integrable point, the kinks remain localized on the trajectories $x = x_0 \pm v t  $ until scattering occurs, which partially destroys the quasiparticles as a consequence of integrability breaking. Remarkably, lightcones with the same slope are observed also in the time-continuous limit, as we show for $\tau = 0.1$. In this case, the chiral nature of the quasiparticles is only evident for kinks between $\ket{0}^{\otimes L}$ and $\ket{01}^{\otimes L/2}$ (upper panels in \cref{fig3}). This is a consequence of the restoration of reflection symmetry as $\tau \to 0$: kinks between the two $Z_2$ states have to spread equally in both directions, and there is no distinction between left and right movers.\\
\indent
A similar light-cone dynamics was first observed in Ref.~\cite{Surace2020}. There, the system was initialized in configurations where a defect is created on a $Z_2$ state by flipping a $\ket{1}$ on a site. In terms of the integrable quasiparticles, a $Z_2$ defect is composed of two kinks (one right and one left mover) at distance zero. In fact, single-kink states do not exist with periodic boundary conditions. However, we argue that they are the elementary excitations that are linked to scarring in the PXP model.
\begin{figure}
\includegraphics[scale=0.58]{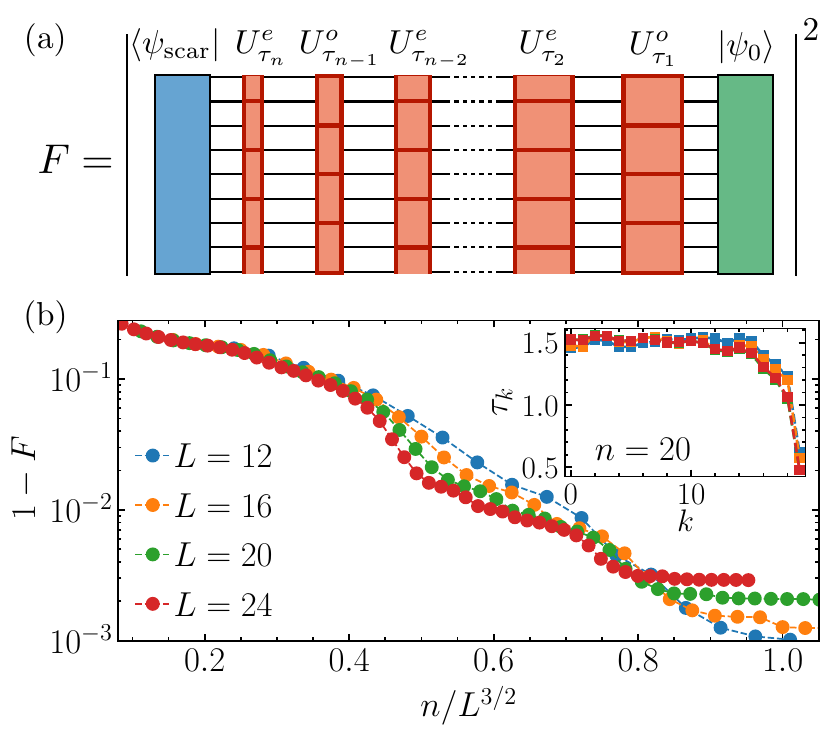}
 \vspace*{-2mm}
\caption{(a) Pictorial representation of the fidelity $F(\{\tau_k\})$ used to obtain the optimal protocol for the preparation of the highest-energy PXP scar $\ket{\psi_\mathrm{scar}}$ starting from the integrable vacuum $\ket{\varepsilon_0 = 2 \pi/3}$. (b) Optimized infidelity as a function of $n/L^{3/2}$ where $n$ is the total number of time steps and for different system sizes $L$. The inset shows the optimal protocol for $n=20$, corresponding to $1-F \simeq 10^{-2}$ for $L=20$. }
\label{fig4}
\end{figure}
\vspace*{2mm}
\paragraph{High-fidelity preparation of PXP many-body scars. --} 
So far, we interpolated between the scar eigenstates of the PXP Hamiltonian \cref{eq:PXP} and few-particle eigenstates of the integrable Floquet operator \cref{eq:floquet} for $\tau  = \pi/2$. We now demonstrate how this interpolation can be reversed to obtain the PXP time-continuous scars starting from the simpler eigenstates of $U_{\pi/2}$. In particular, we engineer a discrete dynamical protocol to prepare the $\ell=1$ PXP scar starting from the $\varepsilon_0 = 2 \pi/3$ vacuum of the integrable model \cref{eq:vacua}. The protocol consists of the successive application of $U^o_\tau$ and $U^e_\tau$ (cf. \cref{eq:floquet}) with a varying time step $\tau_k$ for $k=0,\dots,n-1$, where $n$ is the total number of applied unitaries and the index $k$ plays the role of time in this dynamical protocol. By analogy with dynamical preparation of local Hamiltonian ground states, we expect $\tau_0 = \pi/2$, $\tau_{n-1} \simeq 0$, and a smooth interpolation between these two values for $0 < k < n-1$. Nevertheless, we do not make any assumption on the (discrete) functional form of $\tau_k$ and consider the infidelity $1 - F$ as a cost function (cf. \cref{fig4}(a)), where 
\begin{equation}
\textstyle
F( \{ \tau_k \} ) = \left| \bra{\psi_\mathrm{scar}} \prod_{k=0}^{n/2-1} U^e_{\tau_{2 k+1}} U^o_{\tau_{2k}} \ket{\psi_0} \right|^2.
\label{eq:dyn_prep}
\end{equation}
$\ket{\psi_0} = \ket{\varepsilon_0 = 2 \pi/3}$, and $\ket{\psi_\mathrm{scar}}$ is the time-continuous PXP scar that flows into the integrable vacuum ($\ell = 1$ in \cref{fig2}(a)).
We treat the $\{ \tau_k \}$ as variational parameters, compute the gradient of the cost function, and find the optimal protocol via numerical optimization, starting from the initial condition $\tau_k = \pi/2$ for all $k$ \cite{SupMat}. We show the result in \cref{fig4}(b), where we plot the optimal infidelity as the total number of steps $n$ increases. As expected, the fidelity increases with $n$ for fixed $L$ and decreases with $L$ for fixed $n$. Interestingly, we obtain good data collapse when plotting $1-F$ as a function of $n/L^{3/2}$. The optimal protocol is shown in the inset of \cref{fig4}(b) for $n=20$ and different $L$. We find that the optimal variational parameters are such that $\tau_k \simeq \pi/2$ for most of the process, sharply decreasing only at the end. \\
\indent
The protocol outlined above can be readily implemented in state-of-the-art Rydberg atom platforms, e.g. by making use of two-species arrays to selectively evolve even and odd sites~\cite{bernien}. We refer to \cite{SupMat} for a more detailed discussion of the initial state preparation.

\paragraph{Connections to previous work. --} 
This work is not the first to relate the origin of quantum scars to integrability. 
The possibility that many-body scarring in the PXP model originates from the proximity to an integrable Hamiltonian was first put forward in Ref.~\cite{Khemani2019}. Here, we demonstrated that PXP scars can indeed be smoothly connected to the eigenstates of an integrable model, but realizing a discrete (trotterized) rather than a continuous evolution. As we show in \cite{SupMat}, the Floquet Hamiltonian obtained from a small $\tau$ expansion of the r.h.s.~in \cref{eq:floquet} contains terms that have the same integrability-enhancing effect of the perturbation considered in Ref.~\cite{Khemani2019}. Nevertheless, bridging the PXP Hamiltonian to the integrable CA requires going beyond the perturbative regime at small $\tau$.
A similar approach was used in Ref.~\cite{Rozon2022} where it was shown that the first few terms in the Baker-Campbell-Haussdorf expansion of Eq.~(\ref{eq:floquet}) evaluated at $\tau=\pi/2$ give a small leakage from the subspace spanned by the three vacua. In particular, the fact that the norm of the $n$-th order term decreases with $n$ for small $n$, before eventually growing at larger $n$, was used to argue a possible prethermal regime where the evolution under the PXP Hamiltonian can be approximated with the dynamics of the CA.
Other works \cite{iadecola_ca,Sellapillay2022} focused on the regime $\tau=\pi/2-\epsilon$, showing the persistence of some of the eigenstates of the CA for small perturbations.
Our results show that the two regimes $\tau \rightarrow 0$ and $\tau \rightarrow \pi/2$ are remarkably connected, well beyond the perturbative limit.

\paragraph{Generalizations. --} In the 1D PXP model, $\ket{01}^{\otimes L/2}$ is not the only initial state to yield anomalous dynamics. Another example is the $Z_3$ state $\ket{001}^{\otimes L/3}$, which exhibits revivals with a different period w.r.t. the $Z_2$ state \cite{Turner2018b}. In fact, $\ket{001}^{\otimes L/3}$ has a clear quasiparticles interpretation in the Floquet integrable model as the state with $Q = L/3$ quasiparticles, all separated by two sites. As such, it undergoes a period-5 dynamics under $U_{\pi/2}$. We speculate that the time-continuous anomalous dynamics of this state is also a consequence of the quasiparticles propagation in the integrable PXP CA.
Moreover, other families of PXP scars have been identified \cite{piotr2022}, and we argue that their common property is a large overlap with states that have a small period under $U_{\pi/2}$ \cite{SupMat}, implying that hallmarks of the integrable CA dynamics survive in the time-continuous model beyond the $Z_2$ paradigm.\\ 
\indent
Many-body scars have been observed also in two-dimensional generalizations of the PXP model \cref{eq:PXP} on bipartite lattices \cite{michailidis2020,misha2021}, and are expected to emerge in arbitrary dimension $D$ as well \cite{bennet2022,joey2023}. In fact, the Trotter decomposition \cref{eq:floquet} of the discrete PXP dynamics can be applied to bipartite lattices in any dimension, and, when $\tau = \pi/2$, period-3 oscillations in the subspace spanned by the two $Z_2$ states and $\ket{0}^{\otimes L}$ also occur. However, the interpretation of the full spectrum of $U_{\pi/2}$ in terms of chiral quasiparticles fails in $D>1$. By numerically computing the full spectrum of $U_\tau$ on a square lattice for several system sizes, we found that the linear dependence on $\tau$ of the time-continuous scars eigenvalues is spoiled for $\tau \gtrsim 1$, due to hybridization with the rest of the spectrum \cite{SupMat}. We conclude that our interpretation of PXP quantum scars applies only to the one-dimensional case. 
\vspace*{0.4mm}
\paragraph{Conclusions. --} 
We showed that the celebrated many-body scars of the $1D$ PXP model are smoothly connected to some eigenstates of the integrable PXP CA. The latter are labeled by their number of quasiparticles, and high-temperature PXP scars correspond to the quasiparticles vacua and two-particle eigenstates. 
This correspondence is unraveled by discretizing the PXP time evolution and using the Trotter step as an interpolation parameter. We find that the PXP scar closest to zero (positive) energy descends from one of the quasiparticle vacua, while the following three highest-energy scars are mapped to integrable eigenstates with two quasiparticles. The interpolation does not work equally well for the other scars. Although we attribute this effect to the non-exactness of PXP scars, we could not find a stabilizing perturbation that extends this mapping to lower energy, nor exclude the fact that it might be a finite-size effect. 
Finally, we employed these results to devise a time-discrete dynamical protocol that enables the high-fidelity preparation of the highest-energy PXP scar, opening up new avenues for probing high-temperature scarring eigenstates in experiments.

\let\oldaddcontentsline\addcontentsline
\renewcommand{\addcontentsline}[3]{}
\begin{acknowledgments}
\paragraph{Acknowledgements. --} We acknowledge useful discussions with Kartiek Agarwal, Soonwon Choi, Alessio Lerose, Daniel Mark, Leonardo Mazza, Olexei Motrunich, Silvia Pappalardi, Xhek Turkeshi, and Stefano Veroni.  G.G. acknowledges support from the European Union’s Horizon Europe program under the Marie Sklodowska Curie Action TOPORYD (Grant No. 101106005).  F.M.S. acknowledges support provided by the U.S.\ Department of Energy Office of Science, Office of Advanced Scientific Computing Research, (DE-SC0020290), by Amazon Web Services, AWS Quantum Program, and by the DOE QuantISED program through the theory consortium ``Intersections of QIS and Theoretical Particle Physics'' at Fermilab. H.P. is supported by the ERC Starting grant QARA (Grant No.~101041435), and the European Union’s Horizon 2020 research and innovation program PASQuanS2 (Grant No.~101079862).\\
\end{acknowledgments}

\let\addcontentsline\oldaddcontentsline


\onecolumngrid
\newpage 

\setcounter{equation}{0}
\setcounter{figure}{0}
\renewcommand{\theequation}{S\arabic{equation}}
\renewcommand{\thefigure}{S\arabic{figure}}
\renewcommand{\thesection}{S\arabic{section}}
\renewcommand{\thesubsection}{\thesection.\arabic{subsection}}
\setcounter{secnumdepth}{2}

\begin{center}
    {\Large Supplementary Material \\ 
        \titleinfo
    }
\end{center}
\tableofcontents
	
\section{Few-particle states of the integrable model}
\label{sec:few_particle}
We here show how to compute the eigenstates of the Floquet operator $U_\tau$ for $\tau=\pi/2$, by utilizing the equivalence with the classical CA.
We start by noting that
\begin{align}
    e^{-i\tau P_{j-1}X_j P_{j+1}}&=
    \sum_{n=0}^\infty \frac{1}{n!}\left(-i\tau P_{j-1}X_j P_{j+1}\right)^n\\
    &=1+\sum_{n=1, \text{odd}}^\infty \frac{1}{n!}\left(-i\tau\right)^n P_{j-1}X_j P_{j+1} +\sum_{n=2, \text{even}}^\infty \frac{1}{n!}\left(-i\tau\right)^n P_{j-1}P_{j+1}\\
    &=1-i\sin(\tau)P_{j-1}X_jP_{j+1}+[\cos(\tau)-1]P_{j-1}P_{j+1}.
\end{align}
Therefore, for $\tau=\pi/2$ we get $e^{-i\tau P_{j-1}X_j P_{j+1}}=-i P_{j-1}X_jP_{j+1}+(1-P_{j-1}P_{j+1})$. This operator has the property that it maps classical states (i.e., states of the computational basis) to classical states.

In the following, we will examine some eigenstates of 
\begin{equation}
    U_{\pi/2} =e^{-i(\pi/2) H_e} e^{-i(\pi/2) H_o},
\end{equation}
where
\begin{equation}
    H_{e}=\sum_{j \text{ even}} P_{j-1}X_j P_{j+1},\qquad H_{o}=\sum_{j \text{ odd}} P_{j-1}X_j P_{j+1}.
\end{equation}
We will focus on states within the zero-momentum sectors, i.e., satisfying $T^2=1$.
The general procedure is to start from a classical state $\ket{s}$ and repeatedly apply $U_{\pi/2}$, to obtain a sequence of classical states. As we apply $U_{\pi/2}$, the right and left moving particles of the CA \cite{Prosen2020} propagate ballistically and scatter. After $n\approx O(L)$ steps we obtain the initial configuration $\ket{s}$ up to translations and a global phase. In other words, we have to find the smallest $n>0$ such that the projection of $\ket{s}$ on the zero-momentum sector $\ket{s}_{k=0}$ is an eigenstate of $U_{\pi/2}^n$, with $U_{\pi/2}^n \ket{s}_{k=0}=e^{i\alpha}\ket{s}_{k=0}$. From this parent state, we can construct $n$ different eigenstates $\ket{s, m}$, labeled by $m=0,1,\dots n-1$ of $U_{\pi/2}$. These are defined as
\begin{equation}
\label{eq:eigstates}
    \ket{s, m}\equiv \frac{1}{\sqrt{n}} \left(\sum_{j=0}^{n-1}  e^{-2\pi i mj/n} e^{-i\alpha j/n} U_{\pi/2}^j\right)\ket{s}_{k=0}.
\end{equation}
We can now check that these states are eigenstates of $U_{\pi/2}$ with quasienergies $(\alpha+2\pi m)/n$:
\begin{equation}
    (e^{-2\pi i m/n} e^{-i\alpha/n}U_{\pi/2}-1)\left(\sum_{j=0}^{n-1}  e^{-2\pi i mj/n} e^{-i\alpha j/n} U_{\pi/2}^j\right)\ket{s}_{k=0}=(e^{-i\alpha}U_{\pi/2}^n-1)\ket{s}_{k=0}=0.
\end{equation}

\begin{figure}
    \centering
    \includegraphics[width=0.45\linewidth]{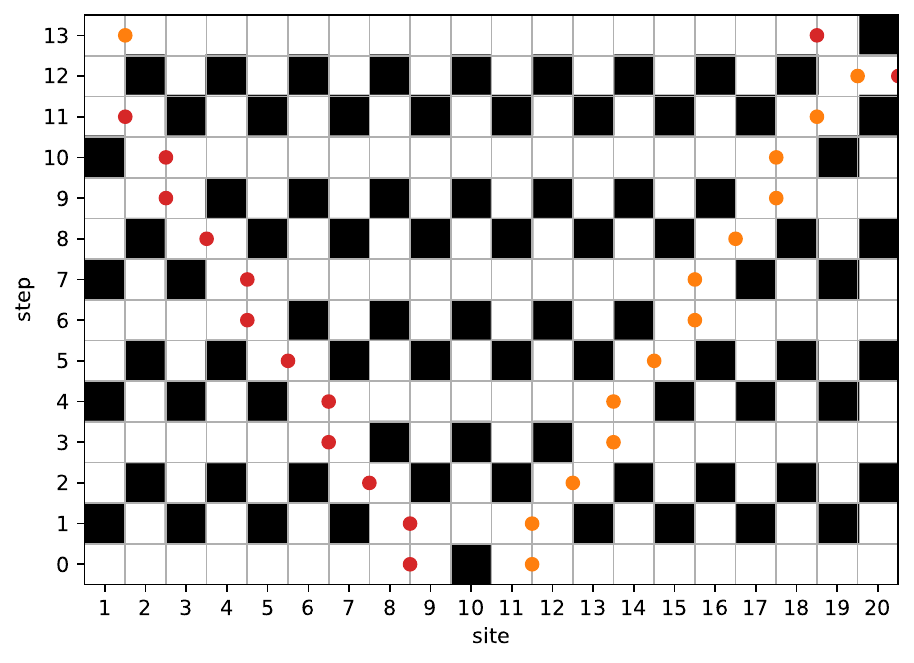}
    \includegraphics[width=0.45\linewidth]{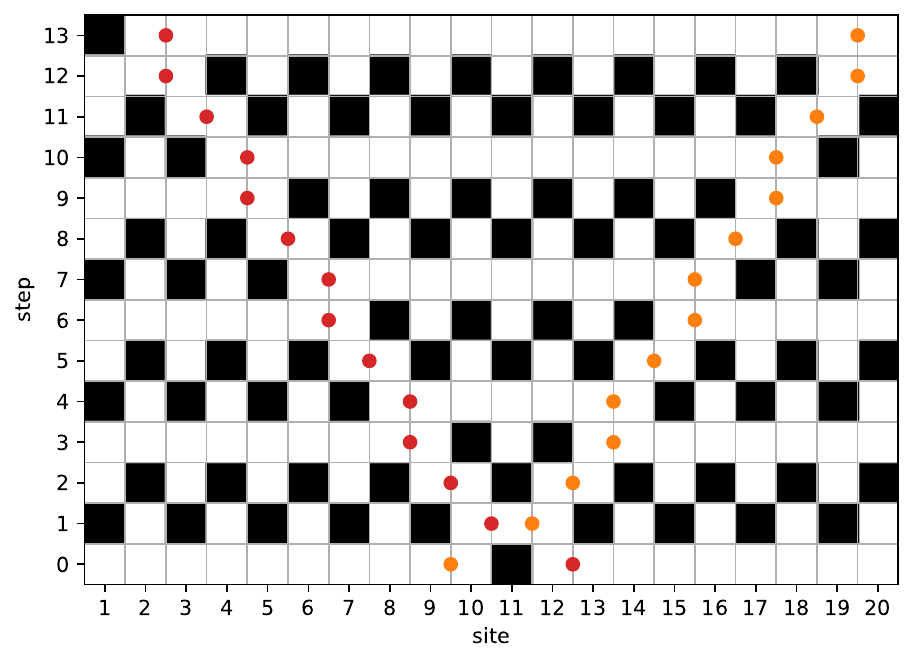}
    \caption{Evolution of the classical configuration with a single $1$ either in an even (left) or odd (right) site, for $L=20$. Red/yellow dots represent left/right moving quasiparticles. The initial configuration (up to a translation of $L/2$ sites) is obtained after $3L/4-2=13$ steps.}
    \label{fig:2part}
\end{figure}

\subsection{Vacua}
Let us consider a system with sites $j=1,\dots L$. We want to construct the vacua states (eigenstates of $U_{\pi/2}$ with no quasi-particles) using the $\ket{000000\dots}$ state as a parent state. We have

\begin{eqnarray}
    U_{\pi/2} \ket{00000000...}&=&(-i)^{L/2}\ket{10101010...},\\
    U_{\pi/2}^2 \ket{00000000...}&=&i^{L/2}\ket{01010101...},\\
    U_{\pi/2}^3 \ket{00000000...}&=&\ket{00000000...}.
\end{eqnarray}

Using Eq.~(\ref{eq:eigstates}) with $n=3$, $\alpha=0$ we find the eigenstates
\begin{align}
    \ket{\varepsilon_0}&\equiv\frac{1}{\sqrt{3}}\left(\ket{00000000...}+e^{- i \varepsilon_0 }U_{\pi/2} \ket{00000000...} +e^{- 2i \varepsilon_0 } U_{\pi/2}^2 \ket{00000000...}\right)\\
    &=\frac{1}{\sqrt{3}}\left(\ket{00000000...}+e^{- i \varepsilon_0 }(-i)^{L/2}\ket{10101010...} +e^{- 2i \varepsilon_0 } i^{L/2}\ket{01010101...}\right)
\end{align}
with quasienergies $\varepsilon_0=0, \pm \frac{2\pi}{3}$.

\subsection{Two-kink states}
Let us focus, for simplicity, on the case $L=0 \text{ mod }4$. We start from a classical configuration with a single $1$, either in an odd or an even site. The evolution under $U_{\pi/2}$ is shown in Fig.~\ref{fig:2part}.
We find that after $n=3L/4-2$ steps we obtain the same configuration of the initial state, up to a translation of $L/2$ sites. It can be shown that for every $L=0 \text{ mod }4$ the global phase after these $3L/4-2$ steps is $e^{i\alpha}=-1$. Therefore, we can use \cref{eq:eigstates} to obtain $n=3L/4-2$ eigenstates with quasienergies 
\begin{equation}
    \varepsilon_m = \pi (1+2m)/n, \qquad m=0,1,\dots n-1.
    \label{eq:kink_energies}
\end{equation}
Since these quasienergies are separated by equal spacings $\Delta \varepsilon= 8\pi/(3L-8)$, we can always find one such quasienergy $\varepsilon$ that satisfies $|\varepsilon-\varepsilon_0|\ge \Delta \varepsilon/2=4\pi/(3L-8)$. These eigenstates having smallest distances from $\varepsilon_0=0, \pm 2\pi/3$ are the $(Q=2)$ states considered in the main text to show a smooth connection to the PXP scar states.
Note that the states obtained by flipping a single $1$ from a CDW state belong to the sequence shown in Fig.~\ref{fig:2part} (at time step $12$ for the left panel, and $1$ for the right panel). These other classical configurations thus generate the same eigenstates. There are in total $2 ( 3L/4 - 2 )$ pairwise degenerate eigenstates with $Q=2$, obtained by applying \cref{eq:eigstates} to the two sets of states depicted in \cref{fig:2part}.

\section{Finite-size scaling of the interpolation from $\tau \to 0$ to $\tau = \pi/2$}
\label{sec:interpolation}

\begin{figure}
    \centering
    \hspace*{-2mm}
     \includegraphics[scale=0.45]{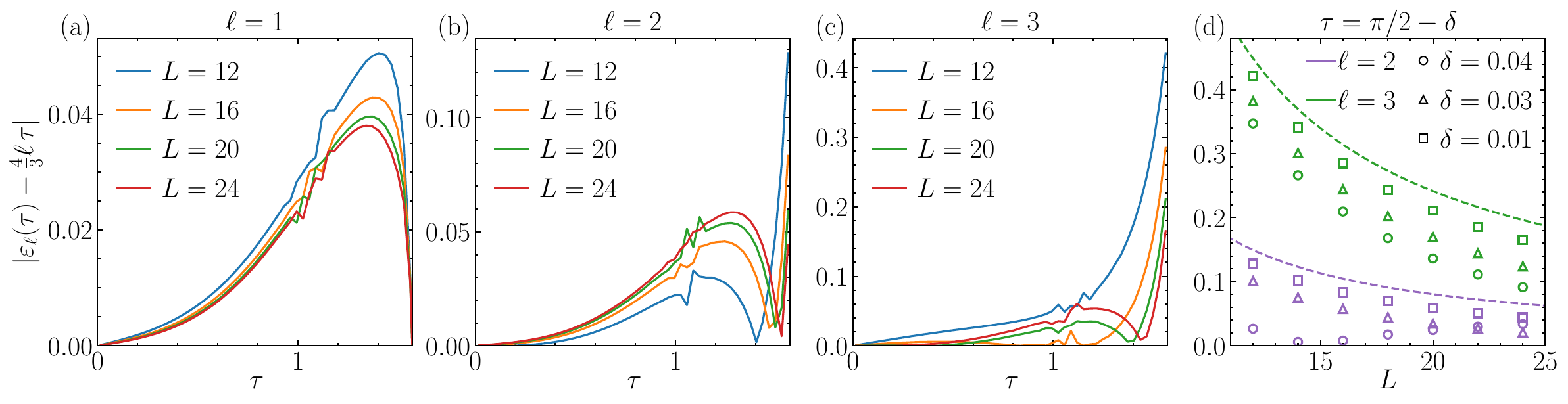}
    \caption{(a-c) Absolute value of the difference between the scars quasienergies $\varepsilon_\ell (\tau)$ and the straight lines $(4/3) \ell \tau$ for $\ell = 1,2,3$ and $L=12,16,20,24$ as a function of $\tau$. (d) Same as (a-c) for $\tau$ close to $\pi/2$. The dashed lines are the two-particle quasienergies of $U_{\pi/2}$ in \cref{eq:kink_energies}.  }
    \label{fig_escaling}
\end{figure}

In the main text, we interpolated between the time-continuous PXP scars with $\ell = 1,2,3,4$ (cf. Fig.~2 of the main text) and the vacua and two-kinks eigenstates of the integral CA $U_{\pi/2}$. In this way, we numerically obtained quasienergies $\varepsilon_\ell(\tau)$ that appear to approach the straight lines $ (4/3) \ell \tau $ as the system size increases. In particular, the quasienergy $\varepsilon_1$ flows directly to the vacuum $\varepsilon_0 = 2 \pi/3$ of $U_{\pi/2}$ and it is almost indistinguishable from the corresponding straight line in Fig.~2 of the main text. In \cref{fig_escaling}(a-c) we show the difference between the $\varepsilon_\ell (\tau)$ and the straight lines $4/3 \ell \tau$ for $\ell=1,2,3$ and different system sizes. For $\ell = 1$ this difference decreases with increasing $L$ uniformly in $\tau$, while for $\ell > 1$ it does not exhibit a clearly decreasing trend. In all cases, we cannot conclude that it eventually vanishes in the thermodynamic limit. In \cref{fig_escaling}(d) we show that, for $\ell > 1$ and $\tau$ close to $\pi/2$, finite-size effects are well captured by \cref{eq:kink_energies} (dashed lines).

\section{Time-discrete dynamical preparation of high-temperature PXP scars}
In this section, we complement the procedure introduced in the main text for the preparation of the highest-energy time-continuous PXP scar ($\ell = 1$ in Fig.~2 of the main text). We first address the preparation of the initial state, i.e., the vacuum of the integrable CA $U_{\pi/2}$ with quasienergy $\varepsilon_0 = 2 \pi/3$, and then provide more details on the numerical optimization employed to obtain the protocol depicted in the inset of Fig.~4 (b) of the main text.\\
We note that a similar preparation protocol can be employed to prepare time-continuous PXP scars for $\ell > 1$, starting from two-kink states of $U_{\pi/2}$. However, as we discussed in \cref{sec:few_particle}, these states are superpositions of $O(L)$ computational basis states, and thus their preparation is unfeasible in experiments. For this reason, we focus on the highest-energy scar $\ell = 1$, which can be obtained starting from the much simpler vacua at $\tau = \pi/2$.  

\subsection{Preparation of the vacua of the integrable model}
\label{sec:prep_vacua}

We focus, for clarity, on the preparation of the state
\begin{equation}
    \ket{\psi_0}= \frac{\ket{0}^{\otimes L }+\ket{01}^{\otimes \frac{L}{2}}+\ket{10}^{\otimes \frac{L}{2}}}{\sqrt{3}},
\end{equation}
as the different vacua states can be straightforwardly obtained from this state (for example, by applying local $Z$ gates via local light shifts). The state $\ket{\psi_0}$ is a generalization of the GHZ state. We will present two protocols for preparing $\ket{\psi_0}$, the first is based on the adiabatic principle, while the second one is based on measurement and feedback.
\subsubsection{Adiabatic preparation}
For this first protocol, we assume that we start with a number of atoms, $M$, that is an integer multiple of 3. We place the atoms equidistantly in a 1D array with next-to-nearest neighbor blockade \cite{Fendley2004,Bernien2017}. For simplicity, we assume periodic boundary conditions, but the discussion generalizes to open boundary conditions if suitable boundary fields are chosen (see. e.g \cite{Omran2019}). The protocol proceeds in four steps:

\begin{enumerate}
    \item 
We start by initializing all atoms in the internal state $\ket{0}$. This is the ground state of the array, when driven at a large negative detuning \cite{Fendley2004,Bernien2017}. 
\item Then the detuning is slowly ramped from negative to positive values, and the Rabi frequency is slowly turned off at the end of this ramp \cite{Fendley2004,Bernien2017}. The state of the Rydberg atom array follows the instantaneous ground state adiabatically, resulting in the preparation of the state $\frac{1}{\sqrt{3}}(\ket{100}^{\otimes M/3}+\ket{010}^{\otimes M/3}+\ket{001}^{\otimes M/3})$. 
\item We now locally address every third atom and drive a resonant, blockade-constrained $\pi$-rotation between the state $\ket{0}$ and $\ket{1}$. This results in the state $\frac{1}{\sqrt{3}}(\ket{100}^{\otimes M/3}+\ket{010}^{\otimes M/3}+\ket{000}^{\otimes M/3})$.
\item Finally, we discard every third atom in the chain. Note that all these atoms are in the state $\ket{0}$, such that the state of the remaining atoms is pure and given by $\frac{1}{\sqrt{3}}(\ket{10}^{\otimes M/3}+\ket{01}^{\otimes M/3}+\ket{00}^{\otimes M/3})$.
\end{enumerate}
In summary, this generates the desired target stated with $L=2M/3$ sites. Note that the $L$ remaining atoms form now a 1D chain with nearest-neighbor blockade constraints.

\subsubsection{Preparation with measurement and feedback}
The second protocol employs measurement and feedback similar to the procedure described in Ref.~\cite{verresen2021efficiently}: this allows us to circumvent the need for a $O(L)$ depth of a unitary circuit, required to prepare long-range entangled states using exclusively unitary evolution.
\begin{figure}[h]
    \centering
    \includegraphics[width=\linewidth]{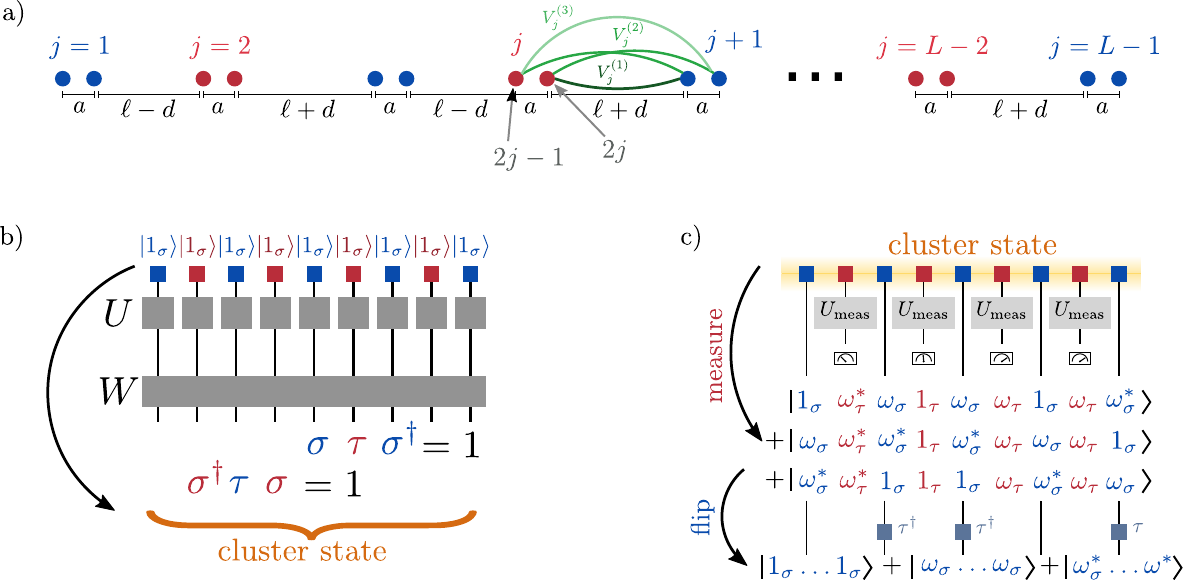}
    \caption{a) $2L-2$ atoms are placed in on a line, with distances $r_{2j,2j+1}=\ell+(-1)^d$ and $r_{2j-1, 2j}=a$. Each pair of atoms $(2j-1,2j)$ represents a clock variable. b) Preparation of a $\mathbb{Z}_3$ cluster state starting from an initial state with no Rydberg excitations. c) Preparation of a $\mathbb{Z}_3$ cat state $\ket{\psi_0}$ from a $\mathbb{Z}_3$ cluster state using measurement and feedback.}
    \label{fig:prep}
\end{figure}
For the preparation of the initial state, we use $2L-2$ atoms. The atoms are displaced from their initial position, such that the distances are $r_{2j,2j+1}=\ell+(-1)^j d$ and $r_{2j-1, 2j}=a $, as in Fig.~\ref{fig:prep}-a. We assume that $a$ is sufficiently small to enforce Rydberg blockade on each pair $2j-1, 2j$, i.e., $n_{2j-1}n_{2j}=0$. For each pair $2j-1, 2j$ we have three possible states $\ket{00}$, $\ket{10}$, $\ket{01}$. We label them as states of a 3-state clock model: 
\begin{equation}
    \ket{1_\sigma}\equiv \ket{00},\qquad \ket{\omega_\sigma}\equiv\ket{10}, \qquad \ket{\omega^*_\sigma}\equiv\ket{01}.\label{eq:clock}
\end{equation}
We define the operators $\sigma, \tau$, which have the following representation in the basis ${\ket{1_\sigma}, \ket{\omega_\sigma}, \ket{\omega^*_\sigma}}$:
\begin{equation}
    \sigma = \begin{pmatrix}
        1 & 0 & 0\\
        0 & \omega & 0\\
        0 & 0 & \omega^*
    \end{pmatrix}, \qquad
    \tau = \begin{pmatrix}
        0 & 0 & 1\\
        1 & 0 & 0\\
        0 & 1 & 0
    \end{pmatrix},
\end{equation}
where $\omega=e^{2\pi i/3}$.
The states defined in Eq.~(\ref{eq:clock}) are eigenstates of the $\sigma$ operator. We will call ${\ket{1_\tau}, \ket{\omega_\tau}, \ket{\omega^*_\tau}}$ the eigenstates of the $\tau$ operator. The $\sigma$ and $\tau$ operators satisfy the relations $\sigma^3=\tau^3=1$ and $\sigma \tau =\omega \tau \sigma$.
The state $\ket{\psi_0}$ is a ferromagnetic $\mathbb{Z}_3$ cat state in the clock variables, i.e., it is an eigenstate of $\sigma_j \sigma_{j+1}^\dagger$ with eigenvalue $1$ for every $j$.

The main idea of the protocol is to prepare a generalized cluster state in the clock variables, which satisfies $\sigma_{j-1} \tau_{j}\sigma_{j+1}^\dagger=1$ for every $j$ even and $\sigma_{j-1}^\dagger\tau_{j}\sigma_{j+1}=1$ for every $j$ odd. Moreover, the $\mathbb{Z}_3$ cluster state satisfies $\prod_j \tau_j=1$. We can then measure all the even-$j$ clock variables in the $\tau$ basis: if the measurement outcome is $\tau_j=1$ we know that $\sigma_{j-1} \sigma_{j+1}^\dagger=1$, as desired; if the measurement outcome is $\tau_j=\omega$ or $\tau_j=\omega^*$ we have one of the two possible types of domain walls. Since we know the position and type of all the domain walls, we can correct them by applying appropriate strings of $\tau$ and $\tau^\dagger$ operators. As a result, the final state for the odd sites will have $\sigma_{j-1}\sigma_{j+1}^\dagger=1$ for $j=2,\dots, L-2$, and will be an eigenstate of $\prod_{j \text{ even}} \tau_j$.

In order for this protocol to work, we need an efficient way of preparing the $\mathbb{Z}_3$ cluster state, of applying the $\tau$ and $\tau^\dagger$ operators, and of measuring in the $\tau$ basis. The first task is more demanding, and we will discuss it below. The $\tau$ operator, on the other hand, can be simply realized by applying $\pi/2$ pulses and local detunings:
\begin{align}
    &\exp\left[-i\frac{\pi}{2}(n_{2j-1}+P_{2j-1}X_{2j})\right]\exp\left[i\frac{\pi}{2}(n_{2j}+P_{2j}X_{2j-1})\right]\\
    &\qquad =(-iP_{2j-1}X_{2j}-in_{2j-1})(iX_{2j-1}P_{2j}+in_{2j})\\
    & \qquad =\Big(\ketbra{10}{00} + \ketbra{01}{10}+ \ketbra{00}{10}\Big)_{2j-1, 2j}=\tau_j.
\end{align}
Applying them in the opposite order gives $\tau_j^\dagger$.

For what concerns the measurement in the $\tau$ basis, it can be realized by applying a series of gates $U_{j,\text{meas}}$ locally on the $j$-th pair of atoms and then measuring the Rydberg occupation. The sequence of gates has the form
\begin{equation}
U_{j,\text{meas}}=e^{i\frac{\pi}{4}X_{2j-1}P_{2j}}e^{i\phi P_{2j-1}X_{2j}}e^{i\frac{\pi}{2}n_{2j-1}}e^{i\frac{\pi}{4}X_{2j-1}P_{2j}}e^{i\frac{5\pi}{6}n_{2j-1}}e^{-i\frac{5\pi}{6}n_{2j}},
\end{equation}
where $\phi=\arccos{(1/\sqrt{3})}$. With a straightforward calculation, it is possible to check that
\begin{equation}
    U_{j,\text{meas}}^\dagger \sigma_j U_{j,\text{meas}}=\tau_j.
\end{equation}

We now discuss the preparation of the $\mathbb{Z}_3$ cluster state. We start from a state with no Rydberg excitations $\ket{00}^{\otimes L/2}=\ket{1_\sigma}^{\otimes L/2}$ and prepare the state $\ket{1_\tau}^{\otimes L/2}$ by applying a uniform Rabi frequency and detuning to all the atoms for a time $T$:
\begin{equation}
    \ket{00}^{\otimes L/2} \rightarrow U\ket{00}^{\otimes L/2} = \exp\left[iT \sum_j (P_{2j-1} X_{2j} + X_{2j-1}P_{2j}+\Delta n_{2j-1}+\Delta n_{2j})\right]\ket{00}^{\otimes L/2} .
\end{equation}
It is convenient to write $U=\otimes_j U_j$ in the $\sigma$ basis as
\begin{equation}
    U_j=\exp\left[iT\begin{pmatrix}
        0 & 1 & 1\\
        1 & \Delta & 0\\
        1 & 0 & \Delta
    \end{pmatrix}\right]_j .
\end{equation}
It is easy to check that by choosing $\Delta=-2$ and $T=\frac{\pi}{2\sqrt{3}}$, we get
\begin{equation}
    U_j \ket{1_\sigma}_j = ie^{-i\frac{\pi}{2\sqrt{3}}}\frac{\ket{1_\sigma}_j + \ket{\omega_\sigma}_j + \ket{\omega^*_\sigma}_j}{\sqrt{3}}=ie^{-i\frac{\pi}{2\sqrt{3}}} \ket{1_\tau}_j,
\end{equation}
which correspond, up to a global phase, to the desired state.

To find a protocol to prepare the $\mathbb{Z}_3$ cluster state, we now make use of a unitary transformation $W$ that satisfies
\begin{equation}
\label{eq:W}
   W\tau_j W^\dagger =\begin{cases}
       \sigma_{j-1} \tau_j \sigma_{j+1}^\dagger & j=2,4,\dots, L-4, L-2,\\
       \sigma_{j-1}^\dagger \tau_j \sigma_{j+1} & j=3,5,\dots, L-5, L-3.
   \end{cases}
\end{equation}
If we apply such transformation to the state $\ket{1_\tau}^{\otimes L/2}$, for every $j$ even we have:
\begin{equation}
    \sigma_{j-1} \tau_j \sigma_{j+1}^\dagger W \ket{1_\tau}^{\otimes L/2}= W\tau_j \ket{1_\tau}^{\otimes L/2}= W\ket{1_\tau}^{\otimes L/2},
\end{equation}
 hence $W\ket{1_\tau}^{\otimes L/2}$ satisfies the defining requirement of the $\mathbb{Z}_3$ cluster state of being an eigenstate of $\sigma_{j-1} \tau_j \sigma_{j+1}^\dagger$ with eigenvalue $1$. We can similarly prove $\sigma_{j-1}^\dagger \tau_j \sigma_{j+1}W\ket{1_\tau}^{\otimes L/2}=W\ket{1_\tau}^{\otimes L/2}$ for $j$ odd.

We now show that the following unitary satisfies Eq.~(\ref{eq:W}):
\begin{equation}
    W = \exp\left[-\frac{2\pi i}{9}\left(\sum_{\ell=1}^{L/2-1}(-\omega\sigma_{2\ell}^\dagger \sigma_{2\ell+1}+\text{H.c.})+\sum_{\ell=1}^{L/2-1}(\omega^*\sigma_{2\ell-1}^\dagger \sigma_{2\ell}+\text{H.c.})\right)\right].
\end{equation}
To prove Eq.~(\ref{eq:W}) for $j$ even (the proof is analogous for $j$ odd), we write
\begin{align}
    W\tau_j W^\dagger &=\exp\left[-\frac{2\pi i}{9}(\omega^*\sigma_{j-1}^\dagger \sigma_{j}-\omega\sigma_j^\dagger \sigma_{j+1}+\text{H.c.})\right]\tau_j\exp\left[\frac{2\pi i}{9}(\omega^*\sigma_{j-1}^\dagger \sigma_{j}-\omega\sigma_j^\dagger \sigma_{j+1}+\text{H.c.})\right] \label{eq:line1}\\
    &=\exp\left[-\frac{2\pi i}{9}\big(\omega^*(1-\omega^*)\sigma_{j-1}^\dagger \sigma_{j}-\omega(1-\omega)\sigma_j^\dagger \sigma_{j+1}+\text{H.c.}\big)\right]\tau_j \label{eq:line2}\\
    &=\exp\left[\frac{2\pi i}{3\sqrt{3}}\big(i \sigma_{j-1}^\dagger \sigma_{j}+\text{H.c.}\big)\right]\exp\left[\frac{2\pi i}{3\sqrt{3}}\big(i\sigma_j^\dagger \sigma_{j+1}+\text{H.c.}\big)\right]\tau_j \label{eq:line3}\\
    &=(\sigma_j^\dagger \sigma_{j-1})(\sigma_j \sigma_{j+1}^\dagger)\tau_j= \sigma_{j-1}\tau_j \sigma_{j+1}^\dagger \label{eq:line4},
\end{align}
where we used $\tau_j \sigma_j = \omega^* \sigma_j \tau_j$ and $\tau_j \sigma_j^\dagger = \omega \sigma_j^\dagger \tau_j$ to go from (\ref{eq:line1}) to (\ref{eq:line2}), and the identity $z=\exp\left[\frac{2\pi }{3\sqrt{3}}(z-z^*)\right]$, valid for $z=1,\omega, \omega^*$, to go from (\ref{eq:line3}) to (\ref{eq:line4}). We further note that $[W, \prod_j \tau_j]=0$, so the state $W\ket{1_\tau}^{\otimes L/2}$ satisfies $\prod_j \tau_j W\ket{1_\tau}^{\otimes L/2}=W\ket{1_\tau}^{\otimes L/2}$, and is therefore the desired cluster state.

We now discuss how to implement the unitary $W$ in the Rydberg atom setup. We note that $W$ is diagonal in the occupation basis of the Rydberg atoms, so we consider the phase that different Rydberg atom configurations acquire when $W$ is applied. We want to implement the same phases by applying the following operator
\begin{equation}
    W_{\text{Ryd}}=\exp\left(i \sum_{j=1}^{L-2} h_{j,j+1}\right),
\end{equation}
\begin{multline}
    h_{j,j+1}=V_j^{(1)} n_{2j}n_{2j+1}+V_j^{(2)} (n_{2j-1}n_{2j+1}+n_{2j}n_{2j+2})+V_j^{(3)} n_{2j-1}n_{2j+2}\\
    +\gamma_j (n_{2j-1}+n_{2j+2})+\delta_j(n_{2j} +n_{2j+1})+ c_j .
\end{multline}
The term $h_{j,j+1}$ contains the interactions between the sites of the $j$-th and $j+1$-th pairs and their local detuning. We now require that
\begin{equation}
    e^{ih_{j,j+1}}=\begin{dcases}
        \exp\left[-\frac{2\pi i}{9}\big(\omega^*\sigma_{j}^\dagger \sigma_{j+1}+\text{H.c.}\big)\right] & j \text{ odd},\\
        \exp\left[\frac{2\pi i}{9}\big(\omega \sigma_{j}^\dagger \sigma_{j+1}+\text{H.c.}\big)\right] & j \text{ even}.\\
    \end{dcases}
    \label{eq:condition}
\end{equation}
\begin{table}[]
    \centering
    \begin{tabular}{c|c|c|c}
         State $(j,j+1)$ & $h_{j,j+1}$ & $\sigma_j^\dagger \sigma_{j+1}$ & condition \\
         \hline
         $\ket{00}\ket{00}$ & $c_j$ & \multirow{3}{*}{$1$} & \multirow{3}{*}{$h_{j,j+1}=\begin{cases}
             +\frac{2\pi}{9} \text{ mod } 2\pi & $j$ \text{ odd}\\
             -\frac{2\pi}{9} \text{ mod } 2\pi & $j$ \text{ even}
         \end{cases}$}\\ 
         $\ket{10}\ket{10}$ & $c_j+\delta_j +\gamma_j +V_j^{(2)}$ & & \\ 
         $\ket{01}\ket{01}$ & $c_j+\delta_j +\gamma_j +V_j^{(2)}$ & & \\ 
         \hline
         $\ket{00}\ket{10}$ & $c_j+\delta_j$ & \multirow{3}{*}{$\omega$} & \multirow{3}{*}{$h_{j,j+1}=\begin{cases}
             -\frac{4\pi}{9} \text{ mod } 2\pi & $j$ \text{ odd}\\
             -\frac{2\pi}{9} \text{ mod } 2\pi & $j$ \text{ even}
         \end{cases}$} \\
         $\ket{10}\ket{01}$ & $c_j+2\gamma_j+V_j^{(3)}$ & &\\
         $\ket{01}\ket{00}$ & $c_j+\delta_j$ & & \\ 
         \hline
         $\ket{00}\ket{01}$ & $c_j+\gamma_j$ & \multirow{3}{*}{$\omega^*$} & \multirow{3}{*}{$h_{j,j+1}=\begin{cases}
             \frac{2\pi}{9} \text{ mod } 2\pi & $j$ \text{ odd}\\
             \frac{4\pi}{9} \text{ mod } 2\pi & $j$ \text{ even}
         \end{cases}$}\\
         $\ket{01}\ket{10}$ & $c_j+2\delta_j +V_j^{(1)}$ & &\\
         $\ket{10}\ket{00}$ & $c_j+\gamma_j$ & &
    \end{tabular}
    \caption{For each configuration of the the clock variables $(j,j+1)$ (first column) we compute $h_{ij}$ (second column) and $\sigma_j^\dagger \sigma_{j+1}$ (third column). For each value of $\sigma_j^\dagger \sigma_{j+1}$, Eq.~(\ref{eq:condition}) gives a condition on $h_{ij}$ (last column). }
    \label{tab:conditions}
\end{table}
For each configuration of the pair of clock variables $(j, j+1)$, we write the condition in Eq.~(\ref{eq:condition}). The conditions are reported in Tab.~\ref{tab:conditions}.
We solve them and we get:
\begin{equation}
    c_j=(-1)^{j-1}\frac{2\pi}{9},
    \hspace{1cm} \delta_j = \begin{cases}
        -\frac{2\pi}{3} & j \text{ odd},\\
        0 & j \text{ even},
    \end{cases}
    \hspace{1cm} \gamma_j = \begin{cases}
        0 & j \text{ odd}\\
        \frac{2\pi}{3} & j \text{ even},
    \end{cases}
\end{equation}
\begin{equation}
    V_j^{(1)} = (-1)^{j}\frac{2\pi}{3} \text{ mod } 2\pi, \hspace{1cm}
    V_j^{(2)}=(-1)^{j-1}\frac{2\pi}{3} \text{ mod } 2\pi,\hspace{1cm}
    V_j^{(3)}=(-1)^{j}\frac{2\pi}{3} \text{ mod } 2\pi.
    \label{eq:VRyd}
\end{equation}
The total local detuning is $\Delta_{2j}=\delta_j+\gamma_{j-1}=0$ for a site in even position $2j$ and $\Delta_{2j-1}= \gamma_j+\delta_{j-1}=0$ for a site in odd position $2j-1$. The only exceptions are the sites close to the boundary, where $\Delta_2=\delta_1=-\frac{2\pi}{3}$ and $\Delta_{2L-2}=\gamma_{L-2}=\frac{2\pi}{3}$.
For what concerns the Rydberg interactions, the conditions in Eq.~(\ref{eq:VRyd}) can be realized with different strategies: for example, the interaction potentials can be tuned by using multiple atomic species, using Rydberg dressing, or displacing along the transversal direction (possibly using anisotropic $p$-state interactions). We will now show that even in the simple linear geometry with a single species, it is possible to approximately achieve the conditions in Eq.~(\ref{eq:VRyd}).
In this case, the interaction terms have the form
\begin{equation}
    V_j^{(1)}=\frac{2\pi C}{[1+(-1)^j x]^6}, \hspace{1cm}
    V_j^{(2)}=\frac{2\pi C}{[1+(-1)^j x+y]^6}, \hspace{1cm}
    V_j^{(3)}=\frac{2\pi C}{[1+(-1)^j x+2y]^6}.
\end{equation}
where we defined $x=d/\ell$, $y=a/\ell$, and $C$ is proportional to the strength of the Rydberg interaction, to the evolution time and to $\ell^{-6}$. By optimizing $C, x, y$ we find the following set of parameters:
\begin{equation}
    \begin{matrix}
        C=1.8525\\
        x=0.0607\\
        y=0.1220\\
    \end{matrix}
    \hspace{1cm}
    \Longrightarrow
    \hspace{1cm}
    \begin{matrix}
    \displaystyle
        V^{(1)}_{\text{even}}=2\pi\left(\frac{4}{3}-0.03\right) & 
    \displaystyle V^{(1)}_{\text{odd}}=2\pi\left(\frac{8}{3}+0.03\right)\\
      
    \displaystyle  V^{(2)}_{\text{even}}=2\pi\left(\frac{2}{3}+0.01\right) & 
    \displaystyle V^{(2)}_{\text{odd}}=2\pi\left(\frac{4}{3}-0.04\right)\\
      
    \displaystyle  V^{(3)}_{\text{even}}=2\pi\left(\frac{1}{3}+0.04\right) & 
    \displaystyle V^{(3)}_{\text{odd}}=2\pi\left(\frac{2}{3}+0.008\right)\\
    \end{matrix}
\end{equation}
We thus found that this choice of parameters locally gives the correct phases, up to corrections of order $(4\cdot 10^{-2}) \times 2\pi$.

\subsection{Numerical details on the protocol optimization}
\label{sec:optimal_control}

To obtain the time-discrete preparation protocol depicted in the inset of Fig.~4 of the main text, we considered the cost function $1-F$, where $F$ is the squared overlap between the target scar state $\ket{\psi_\mathrm{scar}}$ and the final state of the protocol, obtained after applying $n/2$ times the evolution operator $U = U^e U^o$ to the initial state $\ket{\psi_0} = \ket{ \varepsilon_0 = 2 \pi/3 }$ and the final state Eq.~4 of the main text. We use the time steps $\tau_m$ for $m = 0,...,n-1$ as variational parameters for minimizing the cost function. Note that we assume different time steps between even and odd layers. The numerical optimization has been carried out using the method ``L-BFGS-B'' as implemented in SciPy \cite{scipy}, provided with the gradient of the cost function, which reads

\begin{equation}
        \frac{ \partial (1 - F) }{\partial \tau_m } = 
        \begin{cases}
        2 \mathrm{Re} \left[ i \bra{\psi_\mathrm{scar}}  \left( \prod_{k=m/2+1}^{n/2-1} U^e_{\tau_{2 k+1}} U^o_{\tau_{2k}} \right)  H_o  \left( \prod_{k=0}^{m/2} U^e_{\tau_{2 k+1}} U^o_{\tau_{2k}} \right) \ket{\psi_0} \right]  \qquad  & \mathrm{if} \quad m ~ \mathrm{even}, \\[4mm]
        2 \mathrm{Re} \left[ i \bra{\psi_\mathrm{scar}}  \left( \prod_{k=(m+3)/2}^{n/2-1} U^e_{\tau_{2 k+1}} U^o_{\tau_{2k}} \right)  H_e  \left( \prod_{k=0}^{(m+1)/2} U^e_{\tau_{2 k+1}} U^o_{\tau_{2k}} \right) \ket{\psi_0} \right] \qquad & \mathrm{if} \quad m ~ \mathrm{odd} .
        \end{cases} 
\end{equation}
This method allows one to put bounds on the variational parameters $\tau_m$, which we set to the interval $[ 0, 2]$ for all $m$ to ensure stability of the numerical optimization. We use $\tau_m = \pi/2$ for all $m$ as initial condition and observe that different initial conditions result in different protocols, which we found to have a larger cost function at convergence for $n < 100$.

\section{Level statistics of the effective Hamiltonian for small $\tau$}
\label{sec:Hamiltonian}

In this section, we compute the corrections to the PXP Hamiltonian Eq.~(1) of the main text that arise due to the discretized time evolution. As we show below, the second order correction has the same integrability-enhancing effect of the perturbation considered in Ref.~\cite{Khemani2019}. 
We are interested in the Floquet Hamiltonian, i.e., the Hermitian operator $H_\text{eff}(\tau)$ that satisfies
\begin{equation}
    e^{-i\tau H_e} e^{-i\tau H_o}= e^{-i\tau H_{\text{eff}}(\tau)}.
\end{equation}
We remark that the Floquet Hamiltonian is in general non-local in the thermodynamic limit. We can nevertheless consider an expansion for small $\tau$:
\begin{equation}
    H_{\text{eff}}(\tau)\simeq H^{(0)} +\tau H^{(1)}+\tau^2 H^{(2)}+\tau^3 H^{(3)}+\dots,
    \label{eq:expansion}
\end{equation}
where the first terms are
\begin{align}
    &H^{(0)}=H_e+H_o=\sum_j P_{j-1}X_j P_{j+1},\\
    &H^{(1)}=-\frac{i}{2}[H_e, H_o]=-\frac{i}{2}\sum_j (-1)^j P_{j-2} (\sigma_{j-1}^{-} \sigma_{j}^+ -\text{H.c.})P_{j+1}, \label{eq:trotter_1st} \\
    &H^{(2)}=\frac{1}{12}[H_o-H_e, [H_e, H_o]]=-\frac{1}{6}\sum_j ( P_{j-2}\sigma_{j-1}^- \sigma_j^+ \sigma_{j+1}^- P_{j+2}+\text{H.c.})-\frac{1}{12}\sum_j P_{j-1}X_j P_{j+1}(P_{j-2}+P_{j+2}) , \label{eq:trotter_2nd} \\
    &H^{(3)}= -\frac{i}{24}[H_o, [H_e, [H_e, H_o]]].
\end{align}
The second order correction $H^{(2)}$ includes two terms, the second of which is similar to the perturbation considered in Ref.~\cite{Khemani2019}, namely
\begin{equation}
    H_p = \sum_j P_{j-1}X_j P_{j+1}(Z_{j-2}+Z_{j+2}). 
    \label{eq:vedika}
\end{equation}
We introduce the averaged level spacing ratio
\begin{equation}
    r = \overline{ \frac{ \mathrm{min}( E_n - E_{n-1} , E_{n+1} - E_n ) }{ \mathrm{max}( E_n - E_{n-1} , E_{n+1} - E_n)  } },
\end{equation}
where $E_n$ are the eigenvalues of the Hamiltonian and the average is taken over the full spectrum.
In Ref.~\cite{Khemani2019} it was shown that the Hamiltonian $H^{(0)} + h H_p$ is ``almost integrable" for $h\simeq 0.02$, meaning that its level spacing ratio is close to $r \simeq 0.39$, a value expected from the spectrum of an integrable model. On the contrary, the value of $r$ characteristic of a non-integrable model is $r \simeq 0.53$.
To demonstrate that $H_p$ and $H^{(2)}$ are related by the same integrability-enhancing effect on the PXP Hamiltonian, we compute the averaged level statistics $r$ for the following Hamiltonian
\begin{equation}
    H = H^{(0)} + h H_p + 12 g H^{(2)},
    \label{ham3}
\end{equation}
where the factor of $12$ is included to cancel the prefactor in \cref{eq:trotter_2nd}.
In Fig.\ref{fig1}(a) we plot $r$ in the $(h,g)$ plane, which clearly shows a line connecting the points $(h,g) \simeq (0.02,0)$ and $(h,g) \simeq (0,0.02)$ where $r\simeq 0.39$. Interestingly, the value of $g$ for which $r$ is closer to $0.39$ is exactly the same on the two axes, due to the prefactor $12$ in \cref{ham3}. From these results, we argue that the perturbation of Ref.~\cite{Khemani2019} and the second order Trotter term \cref{eq:trotter_2nd} bring the PXP Hamiltonian closer to the same, unknown integrable model.

\begin{figure}
    \centering
    \hspace*{-2mm}
     \includegraphics[scale=0.48]{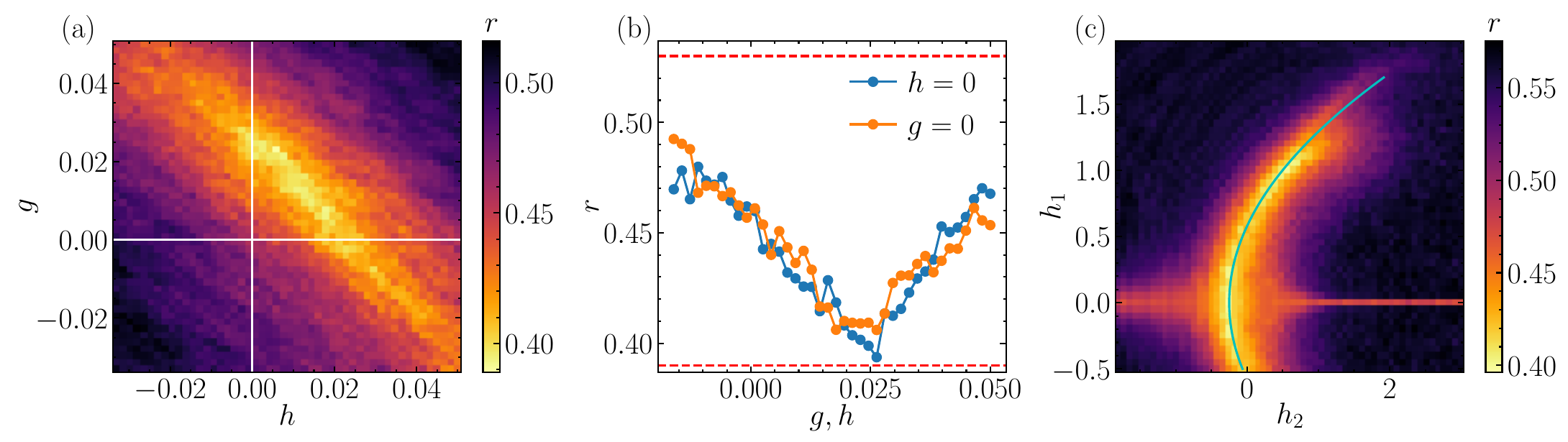}
    \caption{(a) Level spacing statistics for the Hamiltonian Eq.~\eqref{ham3} in the $(g,h)$ plane for $L=24$ sites. (b) Same as (a) for $g=0$ and $h=0$ (white solid lines in panel (a)). (c) Level spacing statistics for the Hamiltonian Eq.~\eqref{ham1} in the $(h_1,h_2)$ plane for $L=24$ sites. The turquoise line is the parabola $h_2 = (3/4) (h_1)^2-0.25$. }
    \label{fig_r}
\end{figure}

The reader might expect this integrability-enhancing effect induced on the PXP Hamiltonian to be related to the fact that $H^{(2)}$ accounts for the time-discretization that eventually leads to the integrable CA $U_{\pi/2}$. However, as we discuss below, this is not the case. 
In fact, upon including $H^{(1)}$ as an additional perturbation to the PXP Hamiltonian we have
\begin{equation}
    H = H^{(0)} + h_1 H^{(1)} + h_2 H^{(2)}.
    \label{ham1}
\end{equation}
We let the two parameters $h_1,h_2$ free to vary and compute the level statistics $r$ in the $(h_1,h_2)$ plane. The result is depicted in \cref{fig_r}(c). $r$ sharply approaches the integrable value $0.39$ on the parabola $h_2 \simeq (3/4) (h_1)^2-0.25$ (solid turquoise line). We now show that this fact has nothing to do with the expansion \cref{eq:expansion}. To see this we compute the first non-trivial correction to the eigenvalues of $H^{(0)}$ induced by the perturbations $H^{(1)}$ and $H^{(2)}$.
The expectation value of $H^{(1)}$ on the unperturbed eigenstates $\ket{n^{(0)}}$ of $H^{(0)}$ vanishes (see the main text). 
Hence the first non-trivial contribution comes at second order and reads
\begin{equation}
\Delta E_n^{(1)} (h_1) = (h_1)^2 \sum_{m\neq n} \frac{\brakett{n^{(0)}}{H^{(1)}}{m^{(0)}} \brakett{m^{(0)}}{H^{(1)}}{n^{(0)}}}{E_n^{(0)}-E_m^{(0)}}.\label{eq:e1}
\end{equation}
The correction induced by $H^{(2)}$ is instead non-trivial at first order and given by 
\begin{equation}
    \Delta E_n^{(2)} (h_2) = h_2 \, \brakett{n^{(0)}}{H^{(2)}}{n^{(0)}} = - \frac{4}{3} h_2  \sum_{m\neq n} \frac{\brakett{n^{(0)}}{H^{(1)}}{m^{(0)}}\brakett{m^{(0)}}{H^{(1)}}{n^{(0)}}}{E_n^{(0)}-E_m^{(0)}}.
    \label{eq:e2}
\end{equation}
To prove the last equality we define $\overline H =H_e-H_o$, and observe that $H^{(1)}=[\overline H, H^{(0)}]/4$ and $H^{(2)}=[\overline H, H^{(1)}]/6$.
Using these expressions, we get
\begin{equation}
\frac{\brakett{m^{(0)}}{H^{(1)}}{n^{(0)}}}{E_n^{(0)}-E_m^{(0)}}=\frac{1}{4} \brakett{m^{(0)}}{\overline H}{n^{(0)}},
\label{eq1}
\end{equation}
which implies
\begin{align}
\brakett{n^{(0)}}{H^{(2)}}{n^{(0)}}&=\frac{1}{6}\sum_m  \left(\brakett{n^{(0)}}{\overline H}{m^{(0)}}\brakett{m^{(0)}}{H^{(1)}}{n^{(0)}}-\brakett{n^{(0)}}{H^{(1)}}{m^{(0)}}\brakett{m^{(0)}}{\overline H}{n^{(0)}}\right) \nonumber\\
&=-\frac{1}{3}\sum_m \brakett{n^{(0)}}{H^{(1)}}{m^{(0)}}\brakett{m^{(0)}}{\overline H}{n^{(0)}},
\label{eq2}
\end{align}
where in the last step we used $\brakett{m^{(0)}}{H^{(1)}}{n^{(0)}}=-\brakett{n^{(0)}}{H^{(1)}}{m^{(0)}}$ and $\braket{n^{(0)}}{\overline H|m^{(0)}}=\brakett{m^{(0)}}{\overline H}{n^{(0)}}$.
Substituting \cref{eq1} into \cref{eq2} we get \cref{eq:e2}. Summing the two non-trivial contributions \cref{eq:e1} and \cref{eq:e2} we finally arrive at
\begin{equation}
    \Delta E_n (h_1,h_2) = \Delta E_n^{(1)} (h_1) + \Delta E_n^{(2)} (h_2) = \left( (h_1)^2 - \frac{4}{3} h_2 \right) \sum_{m\neq n} \frac{\brakett{n^{(0)}}{H^{(1)}}{m^{(0)}} \brakett{m^{(0)}}{H^{(1)}}{n^{(0)}}}{E_n^{(0)}-E_m^{(0)}} 
\end{equation}
From this equation, we see that, to lowest order, 
the spectrum does not separately depend on $h_1$ and $h_2$, but only depends on the difference $(h_1)^2-(4/3)h_2$ and thus it is not surprising that the level statistics is unchanged along 
the parabolas $h_2 = (3/4) (h_1)^2+\text{constant}$. 
Since the point $(h_1,h_2) = (0,0.25)$, which we showed above to be related to the perturbation \cref{eq:vedika} of Ref.~\cite{Khemani2019}, is perturbatively 
small, the result in \cref{fig_r}(c) is explained without resorting to the integrability of the CA $U_{\pi/2}$. On the other hand, the expansion in Eq.~(\ref{eq:expansion}) corresponds to the parabola $h_2=(h_1)^2$, which shows no enhancement of integrability in \cref{fig_r}(c).

 \begin{figure}
    \centering
    \hspace*{-2mm}
     \includegraphics[scale=0.39]{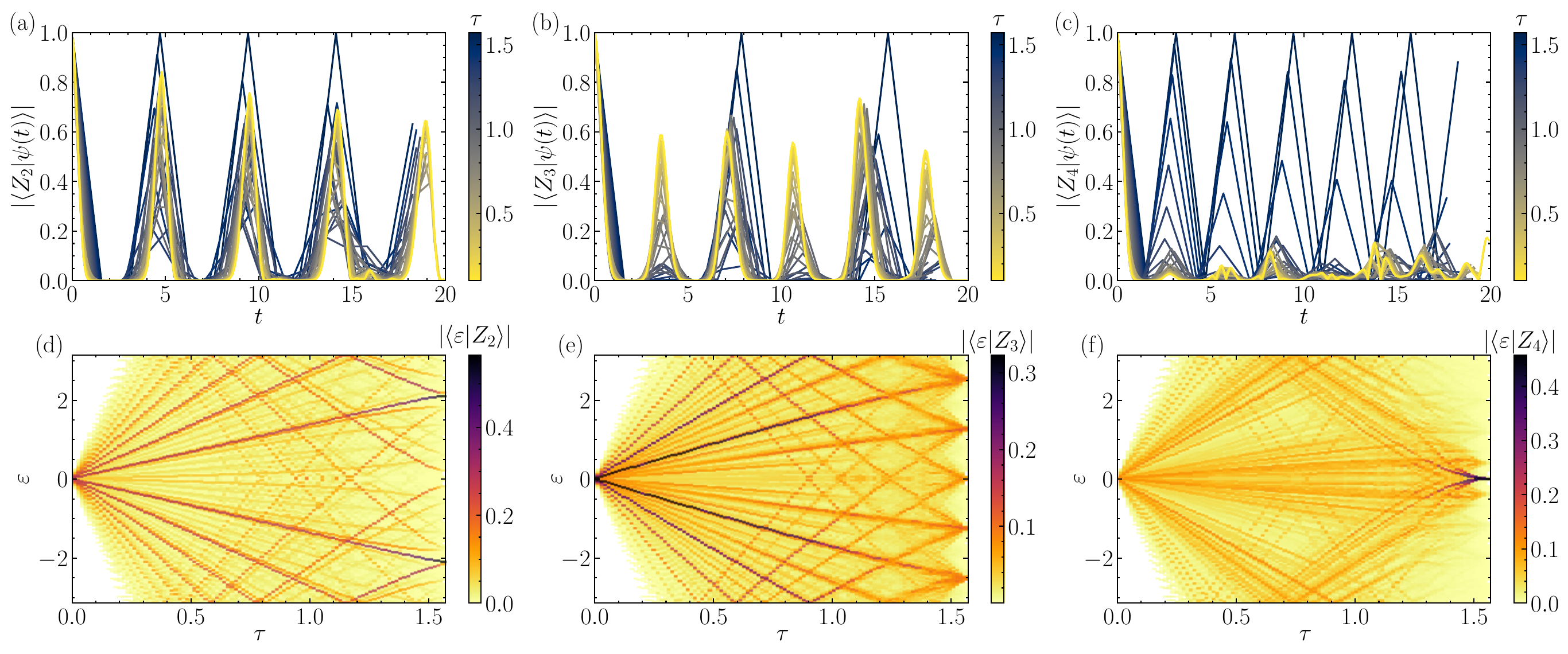}
    \caption{(a-c) Real-time dynamics of the revival fidelity of the $Z_2$,$Z_3$, and $Z_4$ states under $U_\tau$ for $\tau \in [0,\pi/2]$ on a chain of $L=24$ sites. (d-f) Overlap between the eigenstates of $U_\tau$ with $T^2 = T R = 1 $ and the $Z_2$,$Z_3$, and $Z_4$ states as a function of $\tau$ for $L=24$ sites. }
    \label{fig_zn}
\end{figure}

\section{Anomalous dynamics of $Z_n$ states}
\label{sec:Zn}
In the main text, we argued that our results could shed light on the non-ergodic PXP dynamics of initial states other than the $Z_2$ state. In particular, we noticed that the $Z_3$ state, which is known to exhibit atypical oscillating dynamics under the PXP Hamiltonian in $1D$ \cite{Turner2018b}, also has a small period $T=5$ under the integrable CA $U_{\pi/2}$. Following the same line of arguments as for the $Z_2$ case, one would infer that the time-continuous period for this state shall be approximated by $T_{Z_3} \simeq 5 \pi/2$. As we show in \cref{fig_zn}(b), where we plot the revival fidelity in real time of the $Z_3$ state under $U_\tau$ for $\tau$ varying between $0$ and $\pi/2$, $T_{Z_3}$ gives a good approximation of twice the actual $Z_3$ period for $\tau \to 0 $. Although we do not find a straightforward explanation for this period-halving effect, this observation highlights that the connection between PXP time-continuous dynamics and CA dynamics at $\tau = \pi/2$ extends beyond few-particle eigenstates of the CA. In \cref{fig_zn}(e) we plot the $Z_3$ overlap of the eigenstates of $U_\tau$ for $\tau \in [0,\pi/2]$. We see that also in this case $Z_3$ scars exist that interpolate between $\tau \to 0$ and the $5$ eigenstates of $U_{\pi/2}$ that can be built from the period-5 orbit of the $Z_3$ state under $U_{\pi/2}$. Analogously to the $Z_2$ case, their quasienergies are $\varepsilon = 0, \pm 2 \pi/5 , \pm 4 \pi/5 $, namely the 5 eigenvalues to which the scar quasienergies appear to converge for $\tau \to \pi/2$.\\
In fact, the state with the shortest possible period under $U_{\pi/2}$ is the $Z_4$ state $\ket{0001}^{\otimes L/4}$, which has $T=2$. Contrary to the $Z_3$ state this state does not display clear revivals for $\tau \to 0$. Nevertheless, as we show in \cref{fig_zn}(c), a weak revival can be observed at $T_{Z_4} \simeq 2 \pi/2$, supporting the hypothesis that hallmarks of the CA dynamics survive in the time-continuous limit also for states with an extensive number of CA quasiparticles. We note, however, that putative $Z_4$ scars do not strikingly appear in the interpolation from $\tau = \pi/2$ to $\tau \to 0$, potentially justifying the suppressed revivals of this state in the time-continuous limit (cf. \cref{fig_zn}(f)). 

 \begin{figure}[h]
    \centering
    \hspace*{-2mm}
     \includegraphics[scale=0.45]{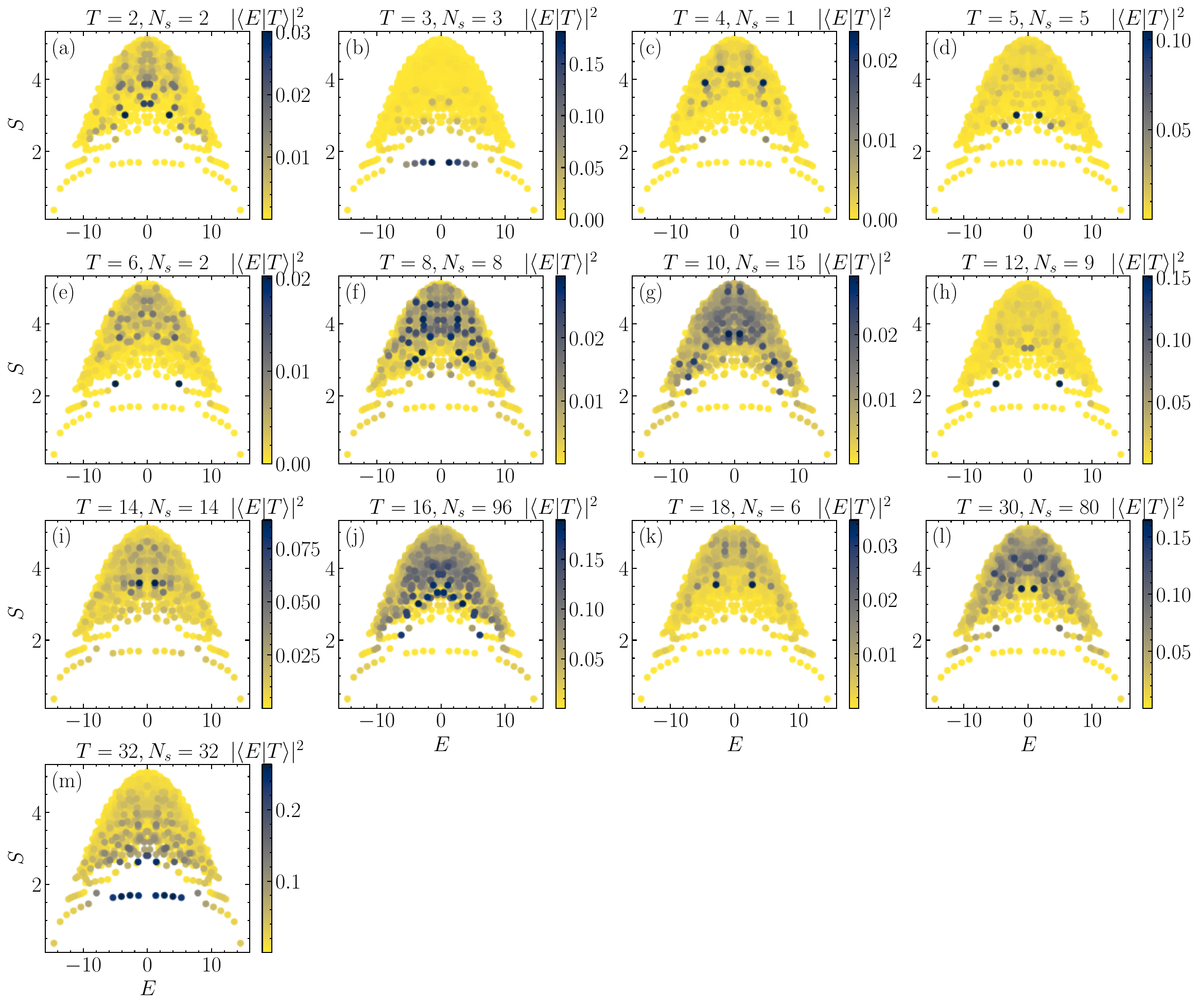}
    \caption{Entanglement entropy of the eigenstates of the time-continuous PXP Hamiltonian as a function of their energy for a chain of $L=24$ sites in the sector where $T^2 = T R = 1 $. In each panel, markers are colored according to their squared overlap on the subspace of states with fixed period $T$ under $U_{\pi/2}$. $N_s$ is the number of states in the fixed-$T$ subspace. The $Z_2$,$Z_3$, and $Z_4$ states correspond to panels $(b)$,$(a)$, and $(d)$, respectively. }
    \label{fig_sub}
\end{figure}

We carried out a more systematic analysis of the connection between states with a short period under $U_{\pi/2}$ and anomalous eigenstates in the time-continuous PXP model. We first select all the subspaces of the Hilbert space that have a period $T \leq 32$ under the classical CA. Among these subspaces, we exclude the ones that have more than $100$ states $N_s$. We then check if these subspaces remain ``localized" onto a few eigenstates of the time-continuous PXP Hamiltonian, and if such eigenstates have low half-chain entanglement entropy, and thus can be considered many-body scars \cite{Turner2018}. To this aim, we plot in \cref{fig_sub} the entropy of the eigenstates of the PXP Hamiltonian as a function of their energy. We color each marker according to their squared overlap on the subspace of period-$T$ states under $U_{\pi/2}$. Remarkably, we see that in several cases the overlap does indeed remain localized onto a few eigenstates, many of which have low entropy despite being in the middle of the spectrum. Therefore, we conclude that the periodic CA dynamics might explain other families of scarring eigenstates in the PXP model, such as the ones identified in Ref.~\cite{piotr2022}.

\section{PXP discrete dynamics on the square lattice}
\label{sec:2D}

\begin{figure}
    \centering
    \hspace*{-2mm}
     \includegraphics[scale=0.79]{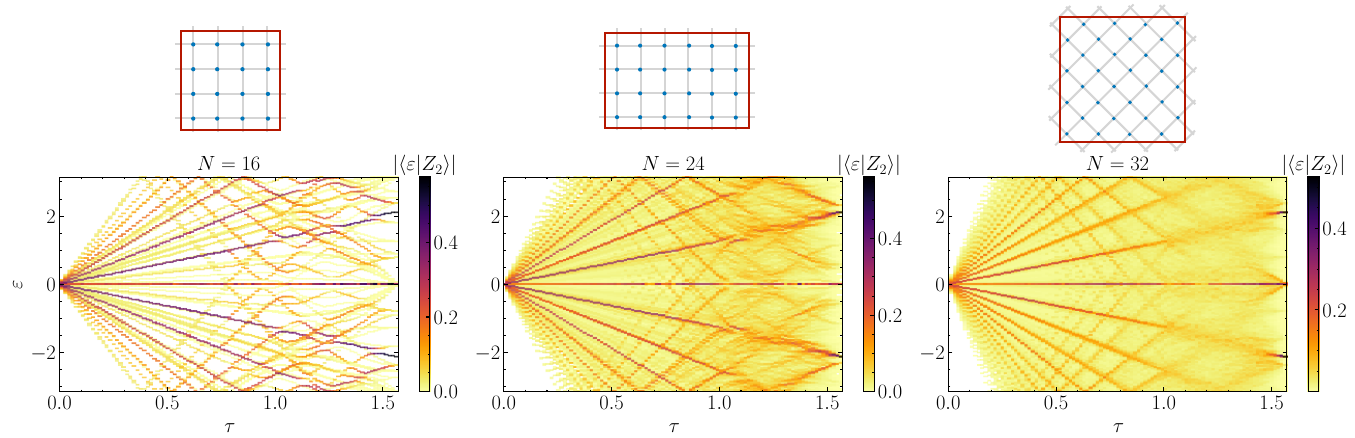}
    \caption{Overlap between the eigenstates of the Floquet operator \cref{eq:floquet_2d} and the $Z_2$ states on the two-dimensional periodic clusters depicted on top of each panel. }
    \label{fig_2d}
\end{figure}

Quantum scarring in the PXP model is expected to persist in higher dimension $D$. In particular, for $D=2$ it has been first studied in Ref.~\cite{michailidis2020}, where it was shown that the two-dimensional case, on bipartite lattices, exhibits a phenomenology analogous to one dimension. Similarly to the $D=1$ case, in $D=2$ scar eigenstates are most easily identified from their large overlap with the $Z_2$ states. In this section, we provide evidence that our findings are instead special to $D=1$. The time-discretization Eq.~2 of the main text can be straightforwardly generalized to any dimension, as long as the lattice is bipartite. Here, we consider a $2D$ square lattice and analyze the spectrum of the Floquet operator
\begin{equation}
    U_\tau = e^{-i H_A \tau } e^{-i H_B \tau },
    \label{eq:floquet_2d}
\end{equation}
where $H_A$ and $H_B$ are the PXP Hamiltonian on the sublattices $A$ and $B$ of the square lattice. For each $\tau$ we compute the overlap of the eigenstates of $U_\tau$ with the $Z_2$ states, on periodic clusters, in the (two-site) translation and reflection invariant sector. The result is presented in \cref{fig_2d}, for increasing number of sites $N$. Also in this setup, we note the presence of eigenstates with large overlap with the $Z_2$ states and quasienergies that are linear functions of $\tau$ for small $\tau$. However, in sharp contrast to $D=1$, these eigenstates hybridize with the rest of the spectrum when $\tau \gtrsim 1$, causing the $Z_2$ overlap to spread at all quasienergies. Only for $\tau \to \pi/2$ the $Z_2$ overlap concentrates at quasienergies $\varepsilon = 0, \pm 2 \pi/3 $, as a consequence of the fact that the vacua of the $1D$ chain are also eigenstates of \cref{eq:floquet_2d}, with the same eigenvalues. We note that this conclusion is in agreement with the observed period of the $Z_2$ oscillations on the $2D$ square lattice, which is larger than in $1D$ ($T_{1D} \simeq 4.78  > T_{2D} \simeq 5.15$), and thus cannot be related to the period-3 dynamics of the vacua of $U_{\pi/2}$. We further note that this result does not contradict other interpretations of PXP scars that are valid in any dimension, but rather implies that PXP scarring has a richer structure in $D=1$ due to the correspondence between scars and integrable eigenstates, and provides a quantitative description of this phenomenon that holds only for the one-dimensional case.


\begin{thebibliography}{37}
\bibitem [{footnote3()}]{footnote3}%
  \BibitemOpen
  \bibinfo {note} {With high-temperature scars we refer to the ones that are
  close to the middle of the spectrum.}\BibitemShut {Stop}%
\bibitem [{footnote4()}]{footnote4}%
  \BibitemOpen
  \bibinfo {note} {To make this procedure more stable we select the scar
  eigenstate at Trotter step $\tau_{k+1}$ by maximizing the sum of squared
  overlaps of the corresponding eigenstates at
  $\tau_k,\tau_{k-1},..,\tau_{k-N}$. We verified that the outcome depends on
  $N$ only for the smaller system size $L=12$, which requires $N>10$ with $d
  \tau = 10^{-2} \pi/2$.}\BibitemShut {Stop}%
\bibitem [{footnote1()}]{footnote1}%
  \BibitemOpen
  \bibinfo {note} {The Floquet Hamiltonian at first order in $\tau$ reads $ i
  \log U_\tau / \tau =$ $ H -\frac{i}{2} [H_e,H_o] \tau + O(\tau^2) $, and for
  the eigenstates $\ket{\varepsilon,\tau}$ of $U_\tau$ for $\tau = 0$ one has
  $\brakett{\varepsilon,0 }{[H_e,H_o]}{\varepsilon, 0 } = \brakett{\varepsilon,
  0 }{[H_e,H]}{\varepsilon,0 } = 0$.}\BibitemShut {Stop}%
\end{thebibliography}
\end{document}